\documentclass[twocolumn,aps,nofootinbib,letterpaper,superscriptaddress,showpacs,floatfix,10pt]{revtex4-1}
\usepackage[utf8]{inputenc}

\usepackage{xcolor}
\usepackage{soul}
\definecolor{blue}{rgb}{0.0, 0.0, 1.0}

\usepackage{amsmath}
\usepackage{graphicx}
\usepackage{subcaption}
\usepackage{float}
\usepackage{booktabs}
\usepackage{multirow}
\usepackage{color}
\usepackage{mathtools}
\renewcommand{\arraystretch}{1.2}
\newcommand{\ba}{\begin{eqnarray}}
\newcommand{\ea}{\end{eqnarray}}
\newcommand{\nn}{\nonumber}
\def \be {\begin{equation}}
\def \ee {\end{equation}}
\def \bea {\begin{eqnarray}}
\def \eea {\end{eqnarray}}

\newcommand{\eq}[1]{(\ref{#1})}

\def\a{\alpha}

\begin{document}
\title{Dark Stars and Gravitational Waves: Topical Review}
\author{Kilar Zhang} 
\email[Email: ]{kilar@shu.edu.cn}
\affiliation{Department of Physics, Shanghai University, Shanghai 200444, Mainland China, Shanghai Key Lab for Astrophysics, Shanghai 200234, Mainland China}

\author{Ling-Wei Luo} 
\email[Email: ]{lwluo@ntnu.edu.tw}
\affiliation{Department of Physics, National Taiwan Normal University, Taipei 116, Taiwan}

\author{Jie-Shiun Tsao} 
\email[Email: ]{tsaojieshiun@gmail.com}
\affiliation{Department of Physics, National Taiwan Normal University, Taipei 116, Taiwan}

\author{Chian-Shu Chen} 
\email[Email: ]{chianshu@gmail.com, Corresponding Author}
\affiliation{Department of Physics, Tamkang University, New Taipei 251, Taiwan}
\affiliation{Physics Division, National Center for Theoretical Sciences,
Taipei 10617, Taiwan}

\author{Feng-Li Lin} 
\email[Email: ]{fengli.lin@gmail.com, Corresponding Author}
\affiliation{Department of Physics, National Taiwan Normal University, Taipei 116, Taiwan}

\begin{abstract}
Motivated by recent observations of compact binary gravitational wave events reported by LIGO/Virgo/KAGRA, we review the basics of dark and hybrid stars\footnote{In this review ``hybrid stars'' means ``dark-matter admixed stars'', though this term usually refers to compact stars that compose of both hadronic and quark matter.} and examine their probabilities as mimickers for black holes and neutron stars. This review aims to survey this exciting topic and offer the necessary tools for the research study at the introductory level. Although called a review, some results are newly derived, such as the equations of state for specific dark star models and the scaling symmetry for the Tidal Love number.
\end{abstract}

\setcounter{footnote}{0}

\maketitle
\tableofcontents
\newpage 

\section{Introduction}
Gravity is the weakest fundamental force in nature. Nevertheless, the gravitational waves (GW) were produced from energy-momentum or matter disturbance, as the theory of General Relativity (GR) predicted{\cite{Einstein1916, Einstein1918}}. These GW propagate throughout spacetime without much being affected by the interstellar regions.  In other words, they preserve most of the information from their sources, according to the feebleness of gravitational force!  After more than half a century's efforts, the first GW event of binary black holes (BBH) coalescence, named GW150914, was observed in the year 2015 by  LIGO Scientific Collaboration and Virgo Collaboration~\cite{LIGOScientific:2016aoc}. 
The penetrability of GW opens a window to detect the binary systems and stochastic GW background for complementary study for astronomy. Also, it provides valuable information on the early universe, such as inflation, primordial black holes (PBH), phase transition, topological defects, etc., for cosmology.

Coincidently, not only GW that weakly interacts with the matter but dark matter (DM), which is composed of about a quarter of the energy density in our universe, also feebly couples with ordinary matter. Currently, three ways to unclose the mystery of DM: direct detection is the observation of light or thermal signals that the nuclei or electron recoils scattering by interstellar DM particles~\cite{PandaX-II:2017hlx, PandaX-4T:2021bab, LUX:2016ggv, PICO:2019vsc, Behnke:2016lsk, XENON:2020kmp}, while indirect detection is to search for the excess of cosmic rays from the core regions of galaxies or cluster systems where is believed to have a highly dense distribution of DM~\cite{Gunn:1978gr, Stecker:1978du, HAWC:2017mfa, IceCube:2012ugg, DAMPE:2017cev, AMS:2016oqu, PAMELA:2013vxg, Fermi-LAT:2009ihh, HESS:2011zpk}. The DM pair annihilations produce these extra cosmic rays.  The final method is collider production, which uses the inverse process of the indirect search and tracing the missing energy or searches for long-live neutral particles~\cite{ATLAS:2022izj, CMS:2014mxa}. 

However, gravity couples universally to all matters, including DM. This opens a new opportunity to detect the nature of DM. It is well known that DM plays a critical role in structure formation according to the density perturbations in the early stage of the universe~\cite{Blumenthal:1984bp}. The standard $\Lambda$CDM (collisionless DM) model simulations predict a large-scale structure consistent with the observations, while some puzzles at small-scale structures demand further study~\cite{Hu:1998kj}. There are several possible solutions to the small-scale problems~\cite{Kaplinghat:2015aga,Cen:1994da,Bringmann:2009vf,Vogelsberger:2015gpr,Foot:2016wvj}; in this review article, we focus on the self-interacting DM (SIDM) scenario~\cite{Tulin:2017ara}.  This is also the nature of the DM we probe.

The progenitors of the stellar objects are formed in interstellar gas clouds. They further evolve into more compact objects through various processes such as accretions, nuclear reactions, gravitational collapses, mergers, etc., among which DM halos are known to provide the gravitational well for such a formation process~~\cite{Renzini:2006je,Hamann:1999az}. If {Standard Model (SM)} particles that consist of about 5\% of the energy density in the universe provide a rich catalog of stellar objects such as black holes (BH), neutron stars (NS), ordinary stars, planets, etc. It is arguable that DM, which has about five times more energy density than the SM particles, might also have similar or even richer possible configurations of stellar objects. However, the lack of self-interaction CDM would make it challenging to form a stellar structure. SIDM, on the other hand, would provide a theoretical ground base for the formation of dark stars (DS). For more detailed reviews on DS, please check \cite{1987thyg.book..199I, Freese_2016, Maselli_2017}.

We will not address the formation mechanism of DS but comment on the potential importance of SIDM to have stellar objects in the dark side of the universe. Instead, we study the stable configurations of DS according to GR for various equations of state (EoS), which correspond to different forms of dark scalar self-interactions. Over one hundred binary-system GW events are observed by the LIGO Scientific, Virgo, and KAGRA Collaborations (LVK) from {O1 to O3} in the past years~\cite{LIGOScientific:2018mvr, LIGOScientific:2021usb, LIGOScientific:2021djp}. 
Some recently observed LIGO/Virgo/KAGRA binary events show the masses of the component compact objects may lie in the regions of low and high mass gaps, which are forbidden for the standard BH formation mechanism. Although the exact values of the mass gaps are uncertain, see \cite{Ertl:2019zks, Belczynski:2020bca} for some discussions, our current knowledge of stellar evolution supports their existence. Thus,  it is quite possible that these mass-gap objects can be regarded as DS and called BH mimickers. More information about these compact objects, such as mass-radius relation and tidal deformability, will depend on the nature of DM and be imprinted in the waveform of GW. Besides, the spin-induced multipole moments and oscillation frequencies, e.g., f-mode frequency, GW absorption, and other information not discussed in this review, will also be encoded in the GW signal.   Thus, one can extract the properties of DM through the data analysis of the mass-gap events.  

\begin{figure*}[]
    \centering
\includegraphics[width=0.8\textwidth]{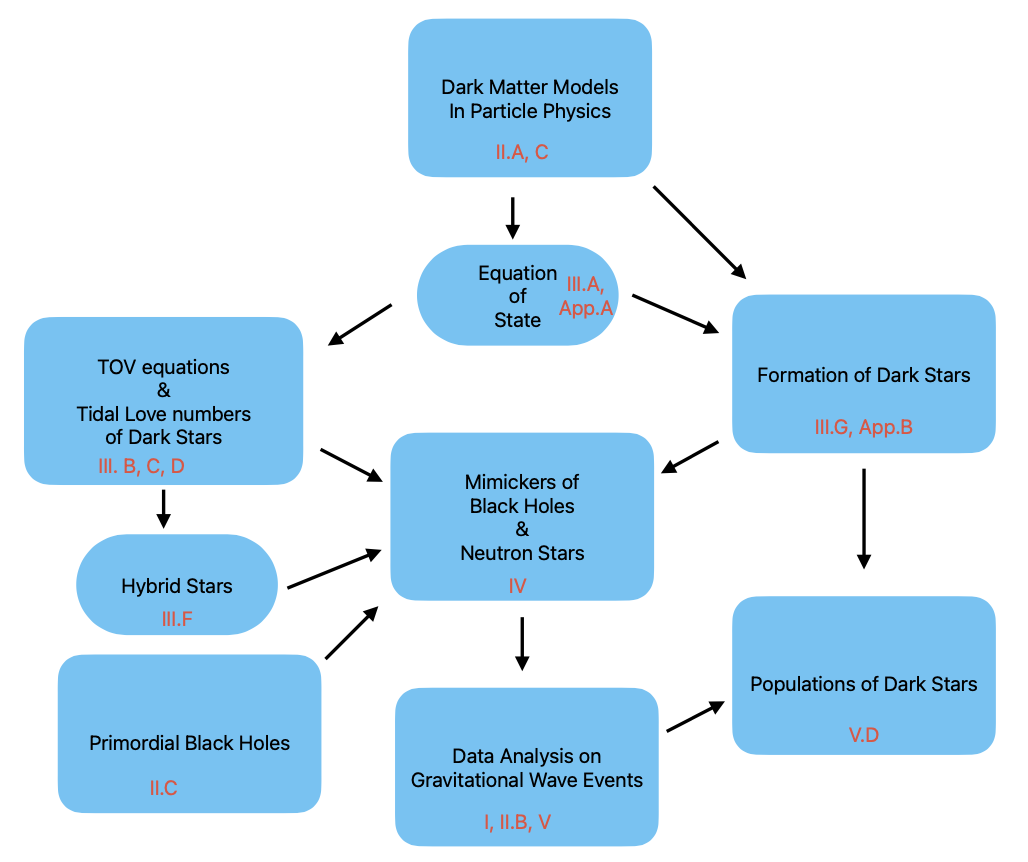}
    \caption{Graphic roadmap of this review with the corresponding sections for each topic indicated in red.}
    \label{fig:outline}
\end{figure*}

As this review covers a broad interdisciplinary landscape across particle physics, astrophysics, and GW astronomy, we  give a graphic outline of this review in Fig. \ref{fig:outline} to orient the readers on how the pieces fit together and provide
an at-a-glance roadmap. 

The plan for the rest of the review goes as follows. The next section reviews the basics of dark energy, DM models, PBH, and their implications for GW physics. Section \ref{dark_hybrid} reviews the models of dark and hybrid stars and their properties, such as stable configurations and tidal deformability, and also sketches the possible formation mechanisms.  Section \ref{mimickers_NB} discusses the boson stars as the mimickers of NS and BH.  Section \ref{data_analysis} briefly reviews the data analysis methodology to infer the existence and properties of DS from GW events. We then conclude our review in section \ref{conclusion}. There are two appendices: the first reviews the derivation of the equation of states for generic bosonic self-interacting DM; the second reviews the Bondi accretion mechanism for star or spike formation out of non-relativistic or relativistic fluids.  In this work, we have adopted $G=c=\hbar=1$ units in most places.

\section{Dark sides of the Universe}
\subsection{Dark energy and dark matter}
It is well established that the energy density of our universe is composed of around 70\% dark energy, 25\% DM, and the rest 5\% is the {SM} particles \cite{PhysRevD.98.030001}. The expansion of our universe is accelerating rather than slowing down by observing the redshift of supernovae \cite{SupernovaSearchTeam:1998fmf}. The negative pressure of dark energy is believed to be responsible for causing the late-time cosmological acceleration. Though the equation of state (EoS) of vacuum energy offers the simplest solution, adding scalar fields also provides a dynamical EoS satisfying the current observations. It would rely on future astrophysical precision measurements to deepen our understanding of this issue. 

Our primary focus, however, in this review article is DM. DM was first proposed for the validity of the virial theorem to infer the stability of cluster galaxies (i.e., Coma galaxy cluster, etc.) and the galaxies (i.e., Milky Way, etc.). A significant component of a missing mass composed by DM is regarded to support the fast infall velocity of satellite objects. Another key reason for introducing the DM is to resolve the rotation curve puzzle, which cannot be explained solely by the visible matter. Similarly, the light bending due to the invisible DM gravitational lens can also explain the novel observed distortion source images.
Furthermore, different epochs in the evolutionary universe, such as the relic abundance of the light nucleus at the early stage and cosmic microwave background (CMB) after the photon decoupling, suggest evidence for the extra energy density of DM.
Last but not least, the primordial energy distributions of DM provide the initial space-time perturbations to the galactic structure formation. The resulting N-body simulation used the initial perturbation condition analyzed from CMB data, consistent with the large-scale survey. 

The best knowledge on the nature of DM is still obscure at the moment of writing; it would be the lack of understanding in gravity theory, or it suggests a new kind of particle. Various experimental searches and theoretical constructions try to solve the mysterious missing piece of energy contribution. In the particle physics aspect, however, it is commonly suggested that DM is presumably a stable particle that interacts inertly with electromagnetic force. Therefore, it is an invisible matter distributed among interstellar space and provides the extra gravitational attraction to sustain the galactic structure. Under this assumption, the nature of DM turns into the quest for its fundamental quantum numbers, such as its spin, its interactions (in terms of Lagrangian), its mass, etc. The form of its interactions with the SM particles will determine the DM relic abundance as the universe evolves. Some mechanisms suggest that DM will thermally freeze out, and some models design for the thermal freeze-in from the thermal bath, to achieve the observed 25\% energy density. It is generally thought that the DM interacts with the SM particles weakly and moves non-relativistically to evolve into the current structure. This is called cold DM (CDM). 

Even though the CDM provides a successful framework to generate large-scale structure, consistent with the observation, some ambiguities at small-scale structure, namely core/cusp problem, missing satellites, and too-big-to-fail, suggest the extension of the CDM model. The collisionless among CDM would generically produce a cuspy density in the central region of the DM halo due to the gravitational accumulation. On the other hand, the observations indicate the core structure is a relatively flat profile. This inconsistency is called the core/cusp problem \cite{deBlok:2009sp, Moore:1994yx, Flores:1994gz, Navarro:1996gj, Walker:2011zu, Oh:2010ea, Maccio:2012qf}. In contrast, the missing satellites problem comes from the number of satellite galaxies surrounding a cluster, which is smaller than the prediction of N-body simulation \cite{Walker:2012td, Boylan-Kolchin:2011qkt, Boylan-Kolchin:2011lmk}. A collisionless CDM is used in the simulation and can produce a DM halo surrounded by several sub-halos. The gravitational potential wells produced by DM halo or sub-halos are believed to be the seeds of galactic structure. Therefore, the numbers of sub-halo suggested in the N-body simulation will be identified as the satellite galaxies. However, the conflict occurs between the simulations and observations. Furthermore, the massive DM halos are often preferred in the N-body simulations that infer that large galaxies should be commonly observed. A measurement of the infall velocity of objects located around the boundaries would tell us the mass content of the galaxy due to the Virial theorem. It is not a regular event to see such a giant galaxy. The name of the problem refers to too-big-to-fail, suggesting this opposite preference. These three small structure puzzles may come from the insufficient understanding of the interaction between DM and SM particles and/or the baryonic processes, such as supernova feedback and photoionization. However, it also suggests the non-trivial feature of DM, such as the DM self-interaction.  Overall, for the proposed DM model to solve the above problems, it imposes the cross-section $\sigma_{\rm DM}$ of self-scattering, and the DM's mass $m_{\rm DM}$ in a small window \cite{Feng:2009hw,vandenAarssen:2012vpm, Tulin:2012wi, Tulin:2013teo, Tulin:2017ara}, 
\be\label{SIDM-constraint}
0.1~{\rm cm^2/g} < \sigma_{\rm DM}/m_{\rm DM} < 1~{\rm cm^2/g}\;.
\ee   
This can serve as a stringent constraint on the DM model with self-interaction.

\subsection{GW signatures and gap events}
Gravity couples universally to stress tensors for all forms of matter, including DM. Therefore, besides the primary goal of testing the strong gravity regime for Einstein's gravity by detecting the GW from distant sources, it also provides a viable venue to detect the sources void of electromagnetic signals, such as the DS.

Since LIGO's first detected BBH event GW150914 \cite{LIGOScientific:2016aoc} found on September 2015, LIGO and Virgo collaborations have finished the third operation run (O3). They will launch their fourth operation run (O4) in the middle of 2023. Up to now, there are about a hundred observed gravitational events of compact binary coalescences (CBC), most of which are BBH events \cite{LIGOScientific:2018mvr, GWTC-2, LIGOScientific:2021djp}. Among them, some are deservedly highlighted discussions. For example, the first discovery of binary neutron stars (BNS) event GW170817 \cite{TheLIGOScientific:2017qsa, Abbott:2018wiz}, which was also detected by the gamma-ray detector \cite{Goldstein:2017mmi} and can shed some light on the equation of state (EoS) of dense nuclear matter. Especially the data analysis of GW170817 shows evidence of a nonvanishing tidal Love number; see \cite{Kastaun:2019bxo} for the more subtle discussions.  Due to the considerable uncertainty of sky location at the current LIGO/Virgo sensitivity stage, detecting the companion electromagnetic signals of BNS or binary neutron star/BH (NSBH) events is usually challenging. For example, the other possible BNS event GW190425 \cite{Abbott:2020uma}, or NSBH events GW200105 and GW200115 \cite{LIGOScientific:2021qlt} do not have the detected electromagnetic follow-up due to the high mass of GW190425 making prompt collapse likely, and the high mass ratios of the two NSBH events making tidal disruption of the NS unlikely.   At least, among the observed GW events up to O3, GW170817 is the only one with detected multi-messenger signals. The later LIGO/Virgo/KAGRA (LVK) operation is hoped to run with enhanced sensitivity, improving the GW events' sky localization and helping find the multi-messenger signals.

The other two LIGO/Virgo events, which fall in the so-called mass gap regimes, are   GW190521 \cite{190521} and GW190814 \cite{LIGOScientific:2020zkf}. The former consists of two BH with masses about $85.3^{+21}_{-14}$ $M_{\odot}$ and $66^{+17}_{-18}$, respectively. The latter consists of a BH with a mass of $22.2-24.3 M_{\odot}$ and a companion compact object with a mass of $2.5-2.67 M_{\odot}$ which is marginally beyond the maximal mass $2.5 M_{\odot}$ of a neutron star \cite{Margalit:2017dij,Rezzolla:2017aly}. 

The mass gap means the range of BH' masses for which the corresponding population of BH is rare \cite{Edelman:2021fik}. There are two mass gaps: 

• The lower mass gap ranges from $2.5$-$5$ $M_{\odot}$ and is mainly supported by the scarcity of the observation events in this range \cite{Kreidberg_2012, Abbott_2019}. However, the physical reason for such a mass gap is unclear. Note that the maximal mass of the NS roughly sets the lower bound. On the other hand, 

• The stage of pair-instability supernovae predicts the upper mass gap of the BH's masses in the stellar evolution, for which the electron-positron production prevents further gravitational collapse and prefers the supernovae explosion. The exact upper mass gap depends on the details of the stellar evolution model \cite{Woosley_2017}, and the one adopted by LIGO/Virgo \cite{Abbott_2019} ranges between $50$ and $150$ solar masses.

• Besides, there is a less mentioned sub-solar mass gap between $0.2$ and $1$ $M_{\odot}$ \cite{Morras:2023jvb} simply because of no viable stellar evolution channel to produce such BH unless invoking the primordial origin due to the density fluctuation at early Universe \cite{Garcia-Bellido:1996mdl}. 

According to LVK's Gravitational-Wave Transient Catalog (GWTC) of compact binary mergers up to O3 \cite{LIGOScientific:2018mvr, GWTC-2, LIGOScientific:2021djp}, there are about a dozen of mass-gap candidate events. Some may have considerable uncertainty on the component masses due to insufficient signal-to-noise (SNR). However, their median values fall in the mass gap. Then, the question is, what are the sources for such mass-gap events? There are three possible scenarios. 

• The first scenario is that they are the secondary objects from previous mergers but not from the collapse of stellar evolution. The observed population tomography and its connection to the formation mechanism can verify or falsify this scenario. One should wait for more observed events to build up the correct tomography. 

• The second one is that they are PBH, which rely on the observed population tomography to be fitted to the inflationary models.

•  The third option is that they are BH mimickers formed by some exotic DM. For a given type of DM, the associated DS will have the specific mass-radius relation as a prediction be verified or falsified by the observed gravitational events. Moreover, as the mimickers of BH, the tidal Love number (TLN) of DS should be tiny to mimic the BH known to have zero TLN. This review will focus on the third option, as it is an exciting interplay between GW astronomy and the mysterious part of particle physics.

\subsection{Dark matter models}
Since there is only astrophysical evidence for DM, which implies that DM mainly interacts with the baryonic matter through gravitational interaction, there is almost no constraint on the DM models as long as it interacts weakly with the standard model particles. Therefore, the weak interacting massive particle (WIMP) model is the simplest DM model. WIMP particles are almost non-interacting and massive enough that they will decouple from the cosmic thermal bath in the early universe. However, it also means that the WIMP is hard to detect by the conventional detector through its interaction with the baryons. The direct search now finds no evidence and severely constrains the masses of WIMP particles and the constant coupling strength with the standard model particles.  

Some natural candidates for the WIMP are neutralino or gravitino of the supersymmetric field theory or supergravity theory~\cite{Hooper:2002nq,Feng:2004mt}. However, as there is no direct evidence for the supersymmetry from the collider experiments, it then puts the physical supports of such candidates with a question mark.  There are other astrophysical phenomena that the WIMP may not explain sufficiently. For example, the core-cusp and missing satellite problems for the dark halos are the discrepancies between the N-body simulations based on the WIMP scenario and the observational structures of small-scale. This then opens the door for alternative DM models. 

Since we focus on the possibility of compact DS for this review,  we will emphasize the DM models compatible with such a possibility.

\subsubsection{Fermionic models}
Up-to-date experimental instruments are still exploring the nature of DM. Depending on the motivation for extending the SM, it could be a fundamental fermion or a bosonic particle. In this subsection, we focus on the fermionic nature of DM particles. Neutrinos were first considered the DM candidate when the neutrino oscillating data did not confirm their mass scales. Though the absolute mass of each SM neutrino is still unknown, the oscillating data and cosmology energy density observation provide the heaviest neutrino to lie around $10^{-1}$~eV to $10^{-2}$-eV~\cite{Capozzi:2013csa,SDSS:2004kqt,Esteban:2018azc,KATRIN:2019yun}. It suggests that the SM neutrinos can not explain the energy density of the missing matter. However, the lightness of neutrino masses is a fundamental question; the famous Type-I seesaw mechanism does provide the right-handed neutrino as a good candidate for DM~\cite{Minkowski:1977sc}. 

If the right-handed neutrinos $\nu_R$ are introduced, their SM quantum numbers are completely neutral. It suggests $\nu_R$ can be the so-called Majorana fermion (a fundamental fermion is an anti-particle of itself), and its mass term is purely a parameter of the theory and not constrained by the SM gauge symmetries. The Lagrangian, which is relevant to neutrino masses, has the form 
\be\label{numassLag}
{\cal L} = {\cal L}_{\rm SM} - Y_{\alpha i}\bar{l}_{\alpha}H\nu_{R i} - \frac{M_{Ri}}{2}\bar{\nu}^c_{Ri}\nu_{Ri} + h.c. 
\ee
where ${\cal L}_{\rm SM}$ is the SM largrangian, $H$ is the SM Higgs doublet, $l_{\alpha}$ are the $SU(2)_{L}$ leptonic doublet with flavor index $\alpha$, $Y_{\alpha i}$ is the Yukawa couplings with $i=1,2,3$ refers to the assumed right-handed species. The number of right-handed neutrinos is not restricted here, and we assume three $\nu_{R}$'s for illustration. Note $c$ is the charged conjugation, and $M_{Ri}$ is the Majorana mass of $\nu_{Ri}$, which is not forbidden by the gauge symmetries, as we mentioned. One does not assume the global lepton number to be necessarily conserved. The Majorana mass matrix $M_{Ri}$ is chosen to be diagonal without loss of generality. After diagonalizing the $6\times6$ neutrino mass matrix in the basis of $(\nu_{e}, \nu_{\mu}, \nu_{\tau}, \nu_{R1}, \nu_{R2}, \nu_{R3})$, one obtains the mass eigenvalues of SM neutrinos as the form 
\be\label{numass}
m_{\nu} = \frac{m_{D}^2}{M_{R}} = \frac{Y^2v^2}{M_{R}}\;.
\ee
Here, we suppress the flavor indices, and $m_D = Yv$ is the Dirac mass of the neutrino, with $v$ being the vacuum expectation value of Higgs. The lightness of neutrino masses can be explained by the suitable choices of the sizes of $Y_{\alpha i}$ and $M_{Ri}$. It was proposed that if one of the right-handed neutrinos has a mass around the keV-scale. It can be a good DM matter candidate and satisfy the current neutrino oscillating data.  

Besides the Type-I seesaw mechanism to generate the small neutrino masses, one may also introduce vector fermions with suitable quantum numbers and impose certain discrete symmetries. The neutrino masses are designed to originate at quantum-loop levels; hence, the corresponding quantity is presumably small. The lightest discrete symmetric odd particle is stable and the candidate for DM. Our argument of fermionic DM in this subsection seems to originate from the neutrino mass mechanism. In general, it is unnecessary; other possibilities, such as mirror fermions~\cite{Hung:2006ap} and the lightest supersymmetric R-parity-odd particles, have their motivations.  

Although we will not discuss this extensively in this review, the fermionic DM with self-interaction or weekly interaction with the visible sector could also be the candidate materials to form DS or hybrid stars. However, the requirements from not destroying the NS \cite{Kouvaris:2011gb, Bramante:2013nma} or from the solar capture \cite{Chen:2014hha} put some constraints on the interaction cross-section, hence on their masses and coupling strengths. Despite that, there is still a wide range of parameter spaces for the fermionic DM to form interesting astrophysical compact objects.  In principle, DM can clump together if the density perturbations satisfy Jean's instability condition or if the dissipative processes due to SIDM would drive the gravothermal evolution. A DS solution is obtained by solving a static and spherically symmetric metric to have the Tolman-Oppenheimer-Volkoff (TOV) equations and combine the specific EoS of the DM model. For details, see, for example, \cite{mukhopadhyay2017compact, Wystub:2021qrn, Lenzi:2022ypb}.  For free fermion, the M-R relation for DS would be scaled by comparing with NS, namely $m_{\rm DM} = 1$~GeV with $M_{\rm DS}\approx1 M_{\odot}$ and $R_{\odot}\approx 10$~km. For other interesting examples in particle physics models, the supersymmetric DM to be around 100 GeV, its DS corresponds to $M_{\rm DS}\approx10^{-4} M_{\odot}$ and $R_{\odot}\approx 10^{-3}$~km. And for the case of right-handed neutrino $m_{\nu_R}=10$~keV, the DS is about $M_{\rm DS}\approx10^{10} M_{\odot}$ and $R_{\odot}\approx 10^{11}$~km. Adding the potential energy due to the SIDM will change the precise M-R relations but not the overall orders, as we provided in the above examples. In \cite{mukhopadhyay2017compact}, the EoS can be abstracted from two-body repulsive interactions, and the fermionic DM admixed NS stability and M-R relations are evaluated for three different cases: static, rigid rotating, and differentially rotating.  It is found that the third case allows the highest mass, with a maximum of up to $1.94$ $M_\odot$ with a radius of about $10.4$ km. Thus, the interacting fermionic DM is also a promising candidate for forming the mimickers for BH and NS.  In \cite{mukhopadhyay2017compact}, the EoS can be abstracted from two-body repulsive interactions, and the fermionic DM admixed NS stability and M-R relations are evaluated for three different cases: static, rigid rotating, and differentially rotating. It is found that the third case allows the highest mass, with a maximum of up to $1.94$ $M_\odot$ with a radius of about $10.4$ km. Thus, the interacting fermionic DM is also a promising candidate for forming the mimickers for BH and NS.

Besides, in \cite{DelGrosso:2023trq}, it is shown that fermion soliton stars also exist at the non-perturbative level, with the solutions numerically found. Then, the whole parameter space of the system is explored, and implications for astrophysical observations/DM searches are deduced. In particular, a standard gas of degenerate neutrons (resp. electrons) can support stable (sub)solar (resp. supermassive) fermion soliton stars with compactness comparable to that of ordinary NS. Thus, fermion soliton stars are compelling neutron star mimickers.

\subsubsection{Bosonic models}

Unlike the fermionic model, there is no degenerate pressure for the bosonic DM model, so it is hard to form compact boson stars for the free massive bosons without the help of degenerate pressure. Surprisingly, it was discovered in \cite{Colpi:1986ye} that compact stars can form by introducing tiny self-interactions. This is because the self-gravitating collapse will rapidly squeeze the boson field into higher density so that the self-interacting repulsive force can balance the gravitational attraction to form compact stars. Indeed, the same effect can also be used to resolve the core-cusp and missing satellites problems of dark halos. In \cite{Colpi:1986ye}, the simplest self-interacting model is considered, namely, the $\frac{\lambda_4}{4} |\phi|^4$ self-interaction for the complex scalar $\phi$ of mass $m$, for which it was argued that a compact star of mass of the order of $\lambda_4^{1/2} M^3_{\rm pl}/m^2$ could form if the following condition holds,
\be\label{phi4limit}
\frac{\lambda_4 M^2_{\rm pl}}{m^2} \gg 1,
\ee 
where $M_{\rm pl}$ is the planck mass. Furthermore, it was also shown in \cite{Colpi:1986ye} that in this limit, the scalar field inside the compact star is in a steady state and can be approximated by a perfect fluid with the following form of the equation of state (EoS) \footnote{ However, in the form of \eq{EoS_Colpi} it is more clear to see how the mass and self-coupling modify the EoS from $\rho=3p$, which is the one for the free massless scalar.},
\be\label{EoS_Colpi}
\rho = 3p + 4 \sqrt{\frac{3m^4 p}{2 \lambda_4}}
\ee 
where $p$ and $\rho$ are the pressure and energy density, respectively. It is then easy to see that this kind of equation of state can yield compact stars by solving the TOV equation as long as \eq{phi4limit} holds.  Besides, due to the simplicity of this model, the cross-section of the self-scattering can be obtained to be $\sigma_{\rm DM}=\frac{9 \lambda^2}{16 \pi m^2}$, by which we can translate the constraint \eq{SIDM-constraint} into the following \cite{Eby:2015hsq}
\be\label{parameter2}
30\left(\frac{m}{\rm GeV}\right)^{3/2} < \lambda_4 < 90\left(\frac{m}{\rm GeV}\right)^{3/2}.
\ee 
Given the mass $m$, this constrains the self-interaction $\lambda_4$ in a very narrow window. This is good for falsifying the model by other constraints, such as the DS candidates from GW events.

Motivated by \cite{Colpi:1986ye}, one can consider other self-interacting bosonic field theories as possible DM candidates, which, most importantly, can also yield compact stars and solve dark halo problems. We now know that the Higgs field is a typical self-interacting scalar. Higher theories for the UV completion of standard model or gravity, such as grand unified theory or string theory, can invoke more scalars with exotic interactions, for example, the dilatons and axons. On the other hand, from the bottom up, we can also have the boson field as the mean field for the Bose-Einsten condensation to yield some superfluid/superconductor states, which can also be the ingredient for the DM and compact boson stars.  Later, we will discuss these possibilities and the associated EoSs and boson star configurations.

\subsubsection{Composite models}
The collisionless CDM encounters difficulty explaining the core density profile, missing satellites, and the too-big-to-fail problems described in previous sections. The idea that DM is composed of new fundamental particles was proposed. This kind of DM is often called Dark nuclei or Dark atoms. The mechanism is to assume the strongly coupled fundamental particles form composite states similar to the quarks form hadrons. In general, due to the hypothetical strong force, the van der Waals force will produce the effects of DM self-interaction. In such cases, the problems of small-scale structure formation can be reconciled. One of the advantages of this scenario is that one may have a series of mass spectra of new composite particles. The variety of composite states could be used to explain DM existence and provide the excess of cosmic rays and DM abundance. We know of large self-interactions among the composite hadrons via the strong nuclear force in the SM. It is, therefore, natural to imagine and investigate the hypothesis that DM-DM self-interactions arise from a new but similar type of composite dynamics. Although suitable parameters and mechanisms to produce the correct relic abundance and the mass spectrum are necessary, in the present Universe, the mass density of DM is about five times larger than that of the SM baryon. This coincidence can be naturally explained when the DM number density has the exact origin as the baryon asymmetry of the Universe, and the DM particle mass is in the GeV range. Such a framework is called asymmetric dark matter (ADM). It is interesting to notice that the mass scale of DM is around the GeV range for various ADM models. In particular, the scarcity of anti-particles in the thermal bath can allow the formation of larger composite bound states like dark nuclei and dark atoms, leading to a very rich phenomenology.  

The typical idea behind such Composite DM models is to provide a stable DM candidate thanks to accidental symmetries in the Lagrangian, similar to proton stability and baryon number conservation in QCD. The visible sector is thus enlarged with a Dark Sector (DS) made of new fermions $\psi$, called dark quarks with quantum numbers of Dark Color based on a certain non-Abelian gauge symmetry such as SU(N) or SO(N). The dark quarks are assumed to be in the fundamental representation of dark color and vector-like representation under the SM. The Lagrangian is given by  
\begin{eqnarray}
L_{DS} = &&- \frac{1}{4}G^{\mu\nu,a}_{D}G_{D\mu\nu,a} + \bar{\psi}_{i}(i\gamma_{\mu}D^{\mu} - m_{\psi})\psi_{i} \nonumber \\
&&+ y_{ij} \psi_{i}\psi_{j}H  + h.c.
\end{eqnarray}
Here, the vector-fermion of dark quark $\psi$ is assumed, and its mass $m_{\psi}$ is a gauge-invariant quantity. The mass spectrum of bound states of $\psi$ is related to the confinement scale $\Lambda_{DS}$ as an analogy to QCD. The cosmological abundance of DM can also be determined by $\Lambda_{DS}$. 

It is also interesting to notice that a new fundamental fermion with QCD $SU(3)_C$ fundamental representation or adjoint representation can form the bound states that satisfy DM's features. The mass scale of such DM lies around 12.5 TeV. 

Here, we illustrate the idea of composite DM by introducing a compelling model named "Quark Nugget Dark Matter." The original idea was based on ref.~\cite{Witten:1984rs} and other proposals afterin~\cite{Parija:1993sq,Lawson:2012vk,Lawson:2012zu,Ge:2019voa}. In this kind of model, the DM is formed by quark and/or antiquark nuggets of huge baryon density (the baryon number is of the order of $|B| > 10^{25}$). The nuggets were produced during the QCD phase transition with a correlation length of the order of the inverse of axion mass ($m^{-1}_a$). At the same time, its stability is protected by the axion domain walls~\cite{Zhitnitsky:2002qa}. An additional feature of this model is the explanation of the matter-antimatter asymmetry via the strong CP axion $\theta$ term, and the abundance of the visible matter to DM matter densities is close to the observed ratio 1:5.  Therefore, in the case of the Quark Nugget model, the fundamental interaction between DM and the ordinary matter is strong interaction rather than the weekly coupled strength. Its abundance is provided by the ratio $\frac{\Omega_{\rm visible}}{\Omega_{DM}}\approx\frac{1}{5}$, while due to their large masses, the number density is small. As a result, this model satisfies current DM direct and indirect observations. The M-R relations of DS composed by the Quark Nuggets are not clear at the moment, and one may resolve the question if the effective theory for the large baryon number fields could be obtained. Finally, one commend to make that our review paper is to study the potential GW signals of DS. An obvious question is the formation of DS, and we found it is generically difficult for the DM fields to provide a good mechanism. The quark nuggets or composite DM models might solve the puzzle.

\subsubsection{Primordial black holes}
The first discussion on the formation of BH in the early universe was given by Zeldovich and Novikov in 1967 \citep{Zeldovich:1967lct}. Independently, Hawking focused on the gravitationally collapsed object with much smaller masses in the early universe in 1971 \citep{Hawking:1971ei}.
Furthermore, the popular model of PBH due to the inhomogeneities of the early Universe was proposed by Carr and Hawking in 1974 \cite{Carr:1974nx}.
Additionally, the idea of PBH regarded as DM was first proposed by Capline in 1975 \cite{Chapline:1975ojl}.

Due to the quantum properties of BH proposed by Hawking, one would be interested in the well-known evaporation effect \cite{Hawking:1974rv, Hawking:1975vcx}. From the discussion, PBH with a mass larger than $10^{15}$g are unaffected by Hawking radiation, and the corresponding lifetime is long enough than the age of the Universe \cite{Hawking:1971ei, Page:1976df}. 
Recall from the constraint of Big Bang nucleosynthesis (BBN); the baryon energy density is at most $5\%$ of critical density \cite{Cyburt:2003fe}. Ordinary BH are formed at late times, all baryonic, and cannot be the candidate for DM. However, PBH are formed in the radiation-dominated era (RD) before BBN and are not constrained by BBN results. Therefore, the non-baryonic property of PBH leads them to cold dark matter (CDM) candidates. PBH mass spectrum and their relic abundance have been estimated in \cite{Carr:1975qj}. In addition, the possible mass windows of PBH have been reviewed in \cite{Carr:2021bzv}. All the up-to-date constraints of PBH are discussed in \cite{Carr:2020gox}.

Below, we give a sketch of the basics of PBH. For more details, the readers can find in recent reviews \cite{Carr:2021bzv, Carr:2020gox, Sasaki:2018dmp, Green:2020jor}.

\paragraph{Formation---} The PBH mass can be estimated through the energy density at the RD. The equation of state can be described in the simple form of $p=w\rho$ with $w=1/3$ at RD. So the energy density and scale factor are given by $\rho\,\propto\, a^{-4}$ and $a\, \propto\, t^{1/2}$. The PBH mass would approximately have an order of horizon mass \cite{Hawking:1971ei, Carr:1974nx}
\begin{align}
M_{\text{PBH}} \sim M_{\text{H}} 
\sim \frac{c^3t}{G}
\sim 10^{15}\,\bigg(\frac{t}{10^{-23} \text{ s}}\bigg)\,\text{g}
\sim  10^{5}\,\bigg(\frac{t}{1 \text{ s}}\bigg)\, M_{\odot}. \label{M_PBH}
\end{align}
So PBH can have mass about $10^{-5}$ g and $10^5\,  M_{\odot}$ if they form at Planck time $t_{\text{pl}} \sim 10^{-43}$ s and $t\sim 1$ s respectively.
There are other formation mechanisms, for example, models of pressure reduction \cite{KHLOPOV1980383, Widerin:1998my, PhysRevD.59.124013}, cosmic string loops \cite{HAWKING1989237, PhysRevD.43.1106, PhysRevD.48.2502, PhysRevD.53.3002, PhysRevD.57.2158}, vacuum bubbles \cite{Crawford:1982yz, PhysRevD.26.2681,Kodama:1982sf, LA1989375,PhysRevD.50.676,1998AstL...24..413K, Khlopov:1999ys}, domain walls \cite{BEREZIN198391, PhysRevD.53.7103, Rubin:2000dq} and string necklaces \cite{Matsuda_2006, Lake_2009}.

The most popular formation model is the gravitational collapse of overdense regions in the early RD universe \cite{Hawking:1971ei}. This mechanism can be easily realized by connecting density perturbation with curvature perturbation. One can consider a homogeneous and isotropic universe described by a spatially flat Robertson–Walker (RW) universe with the scale factor $a(t)$, and its dynamical evolution is governed by the background Friedmann equation
\begin{align}
H^2 = \frac{8\pi G}{3}\bar{\rho},
\end{align}
where $H \coloneqq \dot{a}/a$ with $\dot{a}\equiv da/dt$ and $\bar{\rho}$ is the background energy density. However, we expect a PBH to form at a dense local region, and the energy density could be understood to be perturbed. The locally perturbed region is approximately a spherically symmetric region of positive curvature $K$ and can be described by the metric of the closed universe model, and its dynamical evolution is again governed by the Friedmann equation but with perturbed energy density,
\begin{align}
H^2 = \frac{8\pi G}{3}\rho - \frac{K}{a^2}.
\end{align}
As a result, the density contrast should be 
\begin{align}
\delta \coloneqq \frac{\rho - \bar{\rho}}{\bar{\rho}} = \frac{K}{H^2 a^2}.
\end{align}

The density contrast $\delta$ will evolve up to the order of unity at the time of PBH formation $t_{\text{f}}$, which implies $K/a^2 = 8\pi G\rho/3$. The collapse can be described by
Jeans' instability if the length scale is greater than the Jeans' length
\begin{align}
\lambda_{\text{J}} = \frac{2\pi a}{k_{\text{J}}}
\sim \frac{c_s}{H} = c_s t_{\text{f}}
\end{align}
where the Jeans wavenumber is
\begin{align}
k_{\text{J}} = \frac{a}{c_s}\sqrt{4\pi G \bar{\rho} }
\approx \frac{a}{c_s}H.
\end{align}
Thus, we obtain $\delta(t_{\text{f}}) = \frac{K}{c_s^2 k_{\text{J}}^2}$. Since $\delta(t_{\text{f}})\simeq 1$, this leads to
\be
K \approx c_s^2k_{\text{J}}^2, \qquad {\rm at} \quad t=t_{\text{f}}\;.
\ee

Since the density fluctuation freezes after crossing the horizon, that determines the spectrum of density contrast. The fluctuation with wavenumber $k$ exits the horizon at $t_k$ when 
$k=H(t_k) a(t_k)$. Thus, the density contrast for mode $k$ is 
\be
\delta(t_k)= \frac{K}{H^2(t_k) a^2(t_k)} = \frac{c_s^2k_{\text{J}}^2}{k^2} 
> \frac{c_s^2 k_{\text{J}}^2}{k_{\text{J}}^2} 
= c_s^2.
\ee
Thus, in the RD era, we can obtain the threshold value 
\begin{align}
\delta_c := c_s^2 = \frac{d p}{d \rho} = w = \frac{1}{3}.
\end{align}
\bigskip

\paragraph{Abundance---}
We can write the mass fractions of PBH at the present time $t_0$ and at the formation time $t_{\text{f}}$, respectively, as
\begin{align}
f 
= \frac{\Omega_{\text{PBH},0}}{\Omega_{\text{DM},0}}
\quad\text{and}\quad
\beta = \frac{\Omega_{\text{PBH,f}}}{\Omega_{\text{R,f}}}
= \frac{\Omega_{\text{PBH},0}}{\Omega_{\text{R},0}}a_{\text{f}},
\end{align} 
where $\Omega(z) = \rho(z)/\rho_c$ is the energy density parameter. 
Through the condition of matter-radiation equality $\rho_{\text{R,eq}} = \rho_{\text{DM,eq}}$, or $\Omega_{\text{R},0} = \Omega_{\text{DM},0}\,a_{\text{eq}}$ with $a_{\text{eq}} = a(t_{\text{eq}})$, we can 
a relation of mass fractions of PBH
\begin{align}
f= \frac{a_{\text{eq}}}{a_{\text{f}}}\beta=\frac{1+z_{\text{f}}}{1+z_{\text{eq}}} \beta,
\end{align}
with $z_{\text{eq}} \approx 3500$.  

The size of the PBH formed at $t_{\text{f}}$ will approximately be the contemporary Hubble radius. From the Friedmann equation in the RD era, \begin{align}
H_{\text{f}}^2 = \frac{8\pi G}{3}\bar{\rho}_{\text{f}} = \frac{4\pi^3 G}{45}g_{\ast,\text{f}}\,T^4_{\text{f}}\;,
\end{align}
the Hubble radius at the formation time will be 
\be 
a_{\text{f}}\, \propto\, g^{-1/4}_{\ast,\text{f}}\,T^{-1}_{\text{f}}.
\ee
Moreover, we can also determine the typical mass of PBH, which is the mass contained inside the Hubble horizon, i.e.,
\begin{align}
M_{\text{PBH}} = \gamma M_{\text{H}}(t_{\text{f}}) 
= \gamma \bar{\rho}_{\text{f}}\,\frac{4\pi}{3}H^{-3}_{\text{f}} 
= \frac{\gamma}{2G}H^{-1}_{\text{f}}.
\end{align}
where $\gamma$ is a numerical efficiency factor and depends on the details of gravitational collapse, which can be evaluated as $\gamma \approx 0.2$ \cite{Carr:1975qj}. See also \eq{M_PBH} for the numerical values of the mass span of PBH.

To estimate the initial abundance of PBH at the formation time, the energy density contrast should be larger than its threshold value  $\delta_c$. If the distribution of primordial density perturbations fluctuations is assumed to be a Gaussian distribution 
\begin{align}
P_g(\delta) = \frac{1}{\sqrt{2\pi}\sigma(M_{\text{PBH}})}
\exp\bigg(
\frac{\delta^2}{2\sigma^2(M_{\text{PBH}})}\bigg)
\end{align}
with deviation $\sigma(M_{\text{PBH}})$  obtained by
\begin{align}
\sigma(M_{\text{PBH}}) = \int \frac{dk}{k}\, \mathcal{P}_{\delta}(k)\,W^2(k),
\end{align}
where $\mathcal{P}_{\delta}$ is the power spectrum of the density fluctuation and $W(k)$ is the windows function of the scale $1/(aH)$.
The mass distribution function of PBH can be evaluated according to the Press-Schechter theory \cite{Press:1973iz},
and the result is 
\begin{align}
\beta(M_{\text{PBH}}) = \int^{\infty}_{\delta_c} \gamma P_g(\delta) d\delta 
= \frac{\gamma}{2}\,\mathrm{erfc}\,\nu_c
\approx \frac{\gamma}{2}\frac{e^{-\nu^2_c}}{\sqrt{\pi}\nu_c}.
\end{align}
where we have defined $\nu = \delta/(\sqrt{2}\sigma)$, and the complementary error function $\mathrm{erfc}\, z = 1 - \mathrm{erf}\, z = \frac{2}{\sqrt{\pi}}\int^{\infty}_{z} e^{-t^2} \,dt 
\approx \frac{e^{-z^2}}{\sqrt{\pi}z}$.


In summary, we see that the mass spectrum of PBH covers a wide range. This will contrast the mass spectrum of DS with a limited mass spectrum constrained by the equation of state for DM. 

Moreover, two exciting observations of supermassive BH (SMBH) in the supergiant elliptical galaxy Messier 87 (M87), and in the Milky Way’s center are given by the Event Horizon Telescope (EHT). The formation of SMBH is still a mystery in the field of research. It is concluded by Volonteri that the stellar remnant BH of the accretion is hardly a progenitor of the initial mass of the SMBH \cite{Volonteri:2010wz}. However, the sufficiently large PBH might grow enough by accretion and could still constitute the seeds for the SMBH \cite{Bean:2002kx, Carr:2018rid}. 
Recent observational event of GW190521 from GW by LIGO-Virgo detectors shows the first detection of the intermediate-mass BH (IMBH) \cite{LIGOScientific:2020iuh}. 
The merger event of GW190521 indicates one of the components is more massive than the mass gap of BH, which could be accounted for by a PBH origin.
Of course, these events could also be explained as the black-hole mimickers of some DS, which we will review below.

\section{Dark and hybrid stars}\label{dark_hybrid}
\subsection{Compact stars and equation of state} \label{EoS_sec}

BH are the most compact astrophysical objects, so their mergers can produce detectable strong GW to distant observers. By definition, the compactness $C:=M/r_H$ of a BH is $0.5$ in the $G=c=\hbar=1$ units. To have compact objects other than BH produce detectable GW {by ground-based detectors}, the compactness of such objects should be comparable to the ones of BH\footnote{ {Binary white dwarfs are much less compact but are standard space-based detectors (like eLISA) sources. The scaling of the Newtonian estimate for the merger frequency with compactness and mass is given explicitly in \cite{Giudice:2016zpa}. }}, for example, around $0.1$ to $0.3$. This requires some {repulsive} force of dense matter to counteract the immense gravitational inward force in the late stage of gravitational collapse to avoid the formation of BH. Such an enduring force is so huge that such matter phase is exotic and cannot be found or formed in the Earth's experiments. That is, they can only form in the strong gravity regions of the Universe. Once formed, these matter phases have the exotic equation of states uncommon to daily life. 

One natural origin of such repulsive force is the degenerate pressure of fermions. For example, the electrons' degenerate pressure helps to form the white dwarfs whose $C$'s are about $10^{-3}$. Reaching the compactness comparable to BH requires higher degenerate pressure provided by neutrons or quarks. This is why NS are the most natural candidates as the sources of GW besides BH. However, due to the complication of nuclear theory, such as quantum chromodynamics of describing the neutron or quark fluids, it is hard to derive the EoS of the dense nuclear matter from even the first principle method. Despite that, there are many proposed EoS obtained from alternative or hybrid methods for the dense nuclear matter inside the NS, such as SLy4 \cite{Douchin:2001sv}, Apr4 \cite{Akmal:1998cf} and SKb \cite{Gulminelli:2015csa}.  We may expect this EoS to be pinned down by observing enough GW events of BNS mergers or supernovae, EM observations from NICER, etc. \cite{Dietrich:2020efo}.

On the other hand, the DM with interactions opens avenues to the exotic equations of state. The simplest equation of states from the free fermionic DM of mass $m$ takes the following form \cite{Maselli_2017}
\bea
\rho=\frac{m^4}{8 \pi^2}\big[ x\sqrt{1+x^2}(2x^2+1) -\ln(x+\sqrt{1+x^2}) \big]\;,
\\
p=\frac{m^4}{8 \pi^2}\big[ x\sqrt{1+x^2}(2x^2/3-1) +\ln(x+\sqrt{1+x^2}) \big]
\eea
where $x=k_F/m$ with $k_F$ the Fermi momentum. One can also add various interactions to the fermions to obtain more equations of state. As we will see later, some of these equations of state can yield compactness comparable to one of the BH. 

Besides the fermionic DM, there also exist bosonic ones. Unlike the fermions, the bosons have no degenerate pressure. Therefore, we will not expect to form compact stars from the free bosons. On the other hand, it is natural to expect that the bosonic DM can have self-interaction, as discussed earlier, even though they almost do not interact with standard model particles. Hence, the self-interacting force provides the enduring force against gravitational collapse. Indeed, the self-interactions provide exotic state equations to form the compact boson stars. The first example is proposed in \cite{Colpi:1986ye} for the scalar with the potential $V(\phi)={m^2 \over 2} |\phi|^2 +{\lambda_4 \over 4} |\phi|^4$ theory, which yields the following equation of state in the isotropic limit (${\lambda \over m^2}\gg 1$),
\be\label{EoS_4}
{\rho \over \rho_{\odot}}={3 p\over \rho_{\odot}} + {\cal B}_4 \sqrt{p\over \rho_{\odot}}
\ee
where ${\cal B}_4={0.08 \over \sqrt{\lambda_4}}({m \over \rm{GeV}})^2$ a free parameter. In the above, we have adopted the astrophysical units associated with the solar mass $M_{\odot}$:
\be\label{astrophy}
r_{\odot}=G_N M_{\odot}/c^2,\quad \rho_{\odot}=M_{\odot}/r_{\odot}^3, \quad p_{\odot}= c^2 \rho_{\odot}.
\ee

To follow the same line, one can obtain more exotic equations of state for various self-interacting dark boson models. Below, we give some examples. The first example is the extension of \cite{Colpi:1986ye} with $V(\phi)={m^2 \over 2} |\phi|^2 +{1\over n}{\lambda_n \over \Phi_0^{n-4}} |\phi|^n$. This model has an approximate good UV $Z_n$ symmetry. The corresponding isotropic EoS is
\be
{\rho \over \rho_{\odot}}={n+2 \over n-2}{p\over \rho_{\odot}} + {\cal B}_n ( {p\over \rho_{\odot}})^{2 \over n}
\ee 
where ${\cal B}_n=({2n \over n-2})^{2\over n} ({\rho_{n,0} \over \rho_{\odot}})^{1-{2\over n}}$ with $\rho_{n,0}={m^2 M^2_{\rm pl} \over 4\pi \Lambda_n}$ with $\Lambda_n=(\lambda_n {\Phi_0^2 \over m^2})^{2 \over n-2}{M_{\rm pl}^2 \over \Phi^2_0}$, and $M_{\rm pl}$ the planck mass. The isotropic limit holds when $\Lambda_n \gg 1$.

The second example is the Liouville field with $V(\phi)={m^2 \over 2\beta^2} \Big[e^{\beta^2 |\phi|^2}-1 \Big]$, which is a typical dilaton field in the context of low energy string theory. The corresponding isotropic EoS is parametrized as follows:
\bea 
{\rho \over \rho_{\odot}}&=&{\cal B} \Big(\sigma_*^2 e^{\sigma_*^2} + e^{\sigma_*^2} -1 \Big)\;, \\ 
{p \over \rho_{\odot}}&=&{\cal B} \Big(\sigma_*^2 e^{\sigma_*^2} - e^{\sigma_*^2} -1 \Big)\;,
\eea 
where the free parameter ${\cal B}=\frac{\rho_0}{\rho_\odot}$ with $\rho_0={m^2 M^2_{\rm pl} \over 4\pi \Lambda}$ and $\Lambda=\beta M_{\rm pl}$. The parameter $\sigma_*$ is the scaled $\phi$, and the isotropic limit holds when $\Lambda \gg 1$. From this EoS, we can solve the TOV equation and find that the maximum compactness of the stable stars is 0.194.

The third example is cosh-Gordon field with $V(\phi)= \frac{m^2}{\beta^2}[\cosh(\beta\sqrt{|\phi|^2})-1]$. This model is motivated by the vortex dynamics of the superfluid and can be seen as a kind of superfluid DM. The corresponding EoS in the isotropic limit is 
\bea
{\rho\over \rho_{\odot}} &=& {\cal B} \left(\frac{1}{2}\sigma_*\sinh\sigma_*+ \cosh\sigma_*-1\right) \,,\\
{p\over \rho_{\odot}}&=& {\cal B} \left(\frac{1}{2}\sigma_*\sinh\sigma_*- \cosh\sigma_*+1\right)  \,.
\eea 
The free parameter $\cal B$ is defined as in the case of Liouville field, so are the parameter $\Lambda$ with $\Lambda \gg 1$ the isotropic limit. The maximum compactness of the stable stars from this EoS is 0.182. 

The fourth example is the sine-Gordon field with $V(\phi) = \frac{m^2}{\beta^2}[1-\cos(\beta\sqrt{|\phi|^2})]$. This model is a typical one for the axion field. The corresponding EoS in the isotropic limit is 
\bea
{\rho\over \rho_{\odot}} &=& {\cal B}\left(\frac{1}{2}\sigma_*\sin\sigma_*- \cos\sigma_*+1\right)\,,\\
{p\over \rho_{\odot}}&=& {\cal B}\left(\frac{1}{2}\sigma_*\sin\sigma_*+ \cos\sigma_*-1\right)\,.
\eea 
The free parameter $\cal B$ is defined as in the case of Liouville field, so are the parameter $\Lambda$ with $\Lambda \gg 1$ the isotropic limit. However, this type of EoS has the sinusoidal feature, yielding sensible compact stars only for some range of $\sigma_*$. 

The last example is the one for constructing the non-topological soliton stars \cite{Lee:1991ax, Friedberg:1986tq, Friedberg:1986tp} with $V(\phi) = \frac{1}{2}m^2|\phi|^2(1-\beta^2{|\phi|^2})^2$. The corresponding EoS in the isotropic limit is 
\bea
{\rho\over \rho_{\odot}} &=& {\cal B}\sigma^2_*(1-\sigma_*^2)(1-2\sigma^2_*)\,,\\
{p\over \rho_{\odot}}&=& {\cal B}\sigma^4_*(\sigma_*^2-1)\,.
\eea 
The free parameter $\cal B$ is defined as in the case of Liouville field, so are the parameter $\Lambda$ with $\Lambda \gg 1$ the isotropic limit.

\begin{figure}[H]
    \centering
    \includegraphics[width=0.48\textwidth]{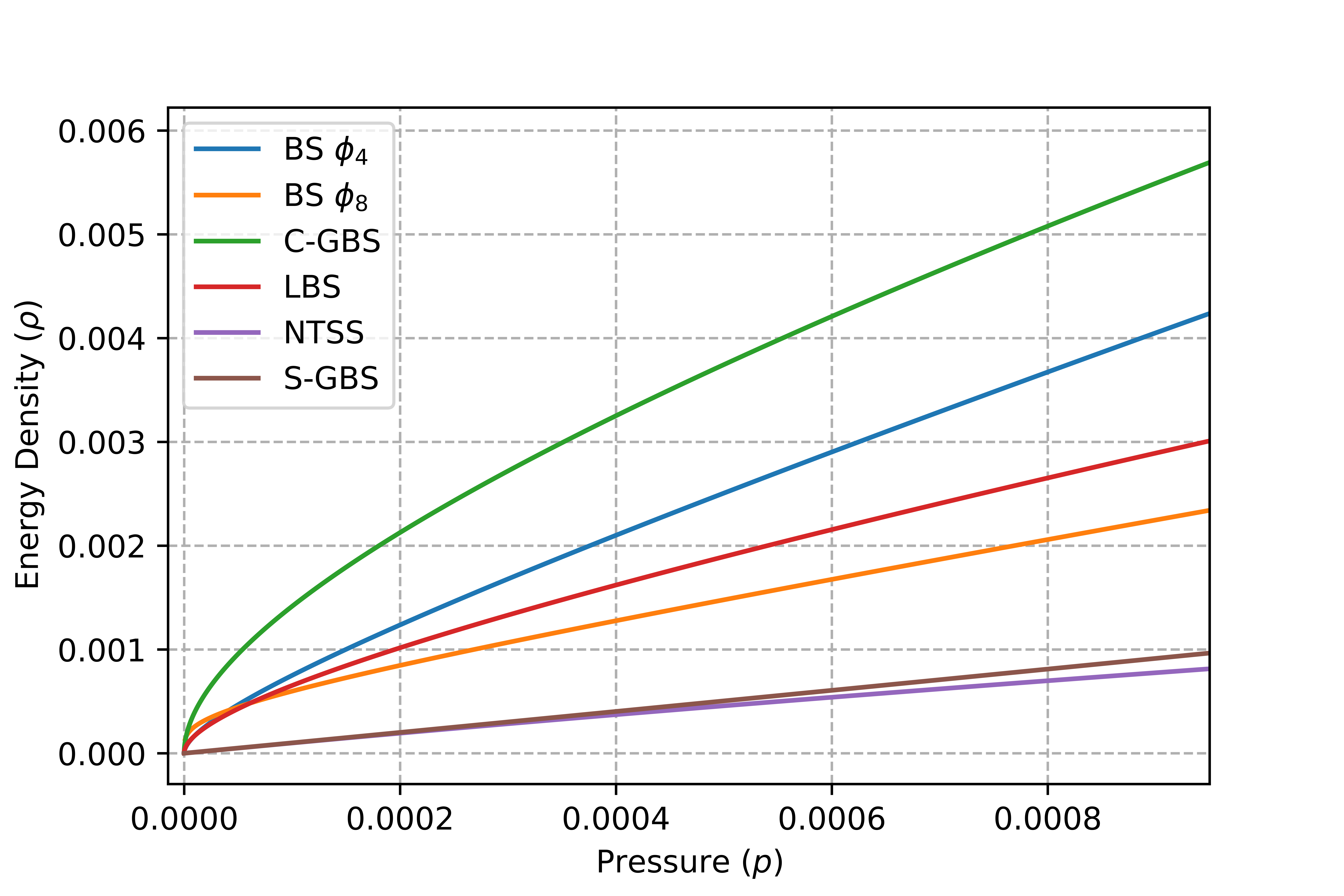}
    \caption{An illustration of the abovementioned EoSs shows their relative behaviors. The adjusting parameters $\mathcal{B}_n$ and $\mathcal{B}$ are chosen to make the maximal masses of the boson stars have several solar masses. We consider $p$ up to $10^{-3}$ since the typical range of the corresponding central pressure is between $10^{-6}$ to $10^{-2}$ as a reference. Here $\rho$ and $p$ are measured in the unit of $\rho_{\odot}$. }
    \label{AllEos}
\end{figure}

In Fig. \ref{AllEos}, we compare the behaviors of the above EoSs.

\subsection{Tolman-Oppenheimer-Volkoff equations}
The EoSs give the relation between $\rho$ and $p$, then the mass and radius of the star can be fixed through the following procedure. 

For a static star with spherical symmetry, consider the Schawarzchild metric 
\small{
\ba\label{metrics}
ds^2=-e^{2\phi(r)} c^2d t^2+\left(1-{2G_N m(r)\over r c^2}\right)^{-1} d r^2 + r^2 d \Omega\;. 
\ea
}
Or, for later convenience, 
\ba\label{metrica}
ds^2=-B(r) d t^2+A(r) d r^2 + r^2 d \Omega\;.
\ea


The Einstein equations read,
\be
G^{\mu \nu}={8\pi \over c^4 }G_N T^{\mu \nu},
\ee
where $G^{\mu \nu}$ can be constructed from the expression of the metric $g^{\mu\nu}$, 
and the energy-momentum tensor is 
\be
T^{\mu \nu}=(\rho+p/c^2) u^{\mu } u^{\nu}+p g^{\mu \nu}.
\ee
Especially in the static case, we can choose $u^{\mu }=(c,0,0,0)$.

Then, from the Einstein equation and the conservation of the energy-momentum tensor, one can obtain the 
 TOV equations \citep{Tolman:1939jz, Oppenheimer:1939ne} , and here we present its multi-component version  \citep{Mukhopadhyay:2016dsg,Rezaei:2018cuk}  with the convention of $G=c=1$:
\be\label{TOV-comp}
{dp_I \over dr}=-(\rho_I + p_I) {d \phi \over dr},\quad  {d m_I \over dr} =4\pi r^2 \rho, \quad  {d \phi \over dr} ={m +4\pi r^3 p \over r(r-2m)},
\ee
where $I$ marks different fluids, $m(r)=\sum_I m_I$ is the total mass inside radius $r$,  total pressure $p=\sum_I p_I$,  and total energy density $\rho=\sum_I \rho_I$ with the contribution from each fluid, and the Newton potential $\phi:={1\over 2} \ln(-g_{tt})$ is introduced in the metric \eqref{metrics}. Taking the single component case, for example, since there are four unknowns as functions of $r$ but with only three equations, we need one more equation to solve them.  In this case, the EoS provides the relation between $p$ and $\rho$.

Given the initial pressure in the center, we can obtain the solution with those equations.   The size  of the star radius  $R$ is taken when $p (r=R)=0$, and the total star mass $M$ is given by $m(R)$.


\subsection{Tidal Love number}
When a static and spherically symmetric star is located under an external quadrupolar tidal field ${\cal E}_{ij}$, it develops a quadrupole moment $Q_{ij}$, both appeared in the metric at large distance $r$ \cite{Thorne:1997kt, Flanagan:2007ix, Hinderer:2007mb},
\begin{eqnarray}
g_{tt}  &=& {-1}+ {2M \over r}+ {3Q_{ij}
\over  r^3} \left( {x^i x^j\over r^2} -
\frac{1}{3}\delta^{ij} \right)+O\left(\frac{1}{r^3}\right)\nonumber\\
& &-{\cal E}_{ij} x^i x^j + O\left(r^3\right),
\end{eqnarray} 
where $M = m(R)$ is the star's total mass.

Then the Tidal Love number (TLN) $\Lambda$ is introduced, defined by
the coefficient to linear order,
\be
Q_{\mu\nu}=- M^5  \Lambda \; {\cal E}_{\mu\nu}\,.
\ee  
Affected by the external tidal field, the metric also suffers a perturbation $h_{\mu\nu}$, and to its linear order, we have
\begin{equation}
g_{\mu\nu}=g^{(0)}_{\mu\nu} +h_{\mu\nu}\,,
\end{equation} 
{where $g^{(0)}_{\mu\nu}$ stands for the unperturbed BH metric.}
Apply the Regge-Wheeler gauge and restrict to the $l=2$, static and even-parity perturbations, $h_{\mu\nu}$ can be expressed as
{\small
\bea
&&h_{\mu\nu} = Y_{2m}(\theta, \varphi) \times\\
&&{\rm diag}
\left[e^{-\nu(r)}H_0(r), ~ e^{\lambda(r)} H_2(r), ~ r^2 K(r), ~
r^2 \sin^2\theta K(r)\right].\nn
\eea}
Then, from the perturbed Einstein equation, we find that $H_2=H_0\equiv H$, which satisfies the differential equation:
\begin{eqnarray}
&&0 = H{''}+H{'} \left[{2 \over r} + e^{\lambda} \left( {2m(r)\over
r^2}
+ 4 \pi r \left(p-\rho\right)\right) \right]  \nn \\
&&+H\left[ -{6 e^{\lambda} \over r^2 } + 4 \pi e^{\lambda}\left( 5
\rho + 9 p +
 {\rho + p \over \left(dp/d\rho\right)} \right)
 - \nu{'}^2 \right]. \qquad  
\end{eqnarray} 
For later convenience, we can introduce
 $y(r):=rH'(r)/H(r)$, then the second order differential equation for $H$ becomes first order: 
\be\label{TLN-y}
ry'+y^2+Py+r^2 Q=0\,,
\ee
where
\bea
P(r)&=&(1+4\pi r^2(p-\rho))/(1-2m/r), \qquad  \\
Q(r)&=&4\pi \frac{5\rho+9p+\sum_I \frac{\rho_I+p_I}{dp_I/d\rho_I}-\frac{6}{4\pi r^2}}{1-2m/r} -4\phi'^2, \qquad \label{multiQ}
\eea
and the boundary condition is now simply $y(0)=2$. Notice that the above equations apply to the multi-fluid case, which is also rigorously derived from the Einstein equation, and the main difference from the single-fluid case is encoded in the $\sum_I \frac{\rho_I+p_I}{dp_I/d\rho_I}$ term of \eq{multiQ}.

Once \eq{TLN-y} is solved, the TLN $\Lambda$ can be obtained from an expression\citep{Hinderer:2007mb,Postnikov:2010yn} of $Y\equiv y(R)$ and the ``compactness" $C=M/R$,
\begin{eqnarray}
&& \Lambda = \frac{16}{15}\left(1-2C\right)^2
\left[2+2C\left(Y-1\right)-Y\right]\times   \nonumber  \\
&&\bigg\{2C\left(6-3 Y+3 C(5Y-8)\right)\\
&& ~ ~+4C^3\left[13-11Y+C(3 Y-2)+2
C^2(1+Y)\right] \nonumber\\
&& ~ ~
+3(1-2C)^2\left[2-Y+2C(Y-1)\right]\log\left(1-2C\right)\bigg\}^{-1}.\nonumber
\end{eqnarray}

\subsection{Scaling symmetry of TOV and TLN configurations}\label{scaling_sec}
By taking a close look into the Mass-Radius curves and TLN-Mass curves for the same series of EoSs, we find that there is a scaling symmetry in the TOV equations and the TLN equation. For the self-similarity in M-R curves, later, we notice that it is already discovered in \cite{Maselli_2017} with an equivalent description, i.e., rewriting TOV into a dimensionless form. While for TLN-Mass curves, the observation is new.

The EoS can usually be described by a pair of parameter functions in the form of 
\bea
{\rho\over \rho_{\odot}} &=& \mathcal{B}\, f(\sigma_*) \,, \label{Ep1}\\
{p\over \rho_{\odot}}&=& \mathcal{B}\, g(\sigma_*)\,,\label{Ep2}
\eea 
where f and g are some arbitrary functions, and $\mathcal{B}$ is a control parameter.
Then we can confirm that, if $\mathcal{B} \to k \mathcal{B}$, then $p \to k p$ is 
consitent with $\rho \to k \rho$.

And it is easy to check that the TOV equation is invariant under the symmetric transformation:
\bea
\rho &\to& k \rho\,,\\
p &\to& k p\,,\\
m &\to & {1\over {\sqrt k}}m\,,\\
r &\to & {1\over {\sqrt k}}r\,.
\eea 
 That is to say, if we set $\mathcal{B} \to k \mathcal{B}$, then the variables change according to the above, while the ``compactness" $C=M/R$ remains the same.

Furthermore, TLN also has the same symmetry. 
From the previous section, it is obvious that $\Lambda$ does not change if we alter $p \to k p$ and $\rho \to k \rho$ simultaneously, since then we have $m \to  {1\over {\sqrt k}}m$ and $r \to  {1\over {\sqrt k}}r$, while $H$ and $ \Lambda $ are homogeneous functions in the order of $m$ and $r$ so that the ${1\over {\sqrt k}}$ scalings will all cancel out.

That is to say, if the EoS can be written in the form of \eqref{Ep1} and \eqref{Ep2} when we alter the parameter $\mathcal{B} \to k \mathcal{B}$, the M-R curves will be similar figures with the ratio of $1\over {\sqrt k}$. In contrast, the TLN-M curves will be magnified by $1\over {\sqrt k}$ only the $M$ axis alone. The maximum compactness and the minimum TLN will remain unchanged when changing $\mathcal{B}$.

We must emphasize that one analytic EoS is generally easy to rewrite in the form of \eqref{Ep1} and \eqref{Ep2}. For example, if we consider a polytropic EoS $\rho=\alpha p^{\gamma}$, then it is equivalent to set $f(\sigma_*)=\sigma_*$, $g(\sigma_*)=\sigma_*^\gamma$ and $\mathcal{B}=\a^{1\over {1-\gamma}}$.

\subsection{I-Love-Q relation}
A neutron star (NS) or quark star (Q) is characterized by macroscopic quantities such as mass $M$, spin angular momentum $J$, angular velocity $\Omega$, the moment of inertia $I=J/\Omega$, quadrupole moment $Q$ and TLN $\Lambda$ (or their dimensionless counterparts  $\bar{I}=I/M^3$,  $\bar{Q}=-Q/(MJ^2)$,  and $\bar{\Lambda}=\Lambda/M^5$, respectively). Since these macroscopic quantities are self-consistently determined from the dynamical equations with a definite EoS, they should depend strongly on the EoS. Surprisingly, it is observed \cite{Yagi:2013bca} that some of these quantities obey universal relations, which are not sensitive to the EoS. They relate the reduced quantities $\bar{I}$, $\bar{\Lambda}$ and $\bar{Q}$, and are named as  \emph{I-Love-Q} relations. 


In detail, taking any two of those above three reduced quantities and denoting them as $y_i$ and $x_i$, the \emph{I-Love-Q} relations take the following form on a log-log scale, 
\begin{equation}
\label{log fit}
\ln y_i = a_i + b_i \ln x_i + c_i (\ln x_i)^2 + d_i (\ln x_i)^3+ e_i (\ln x_i)^4,
\end{equation}
with all the coefficients shown in Table I, which barely varied while changing the EoS. For NS, those coefficients are fitted using six phenomenological EoS including APR \cite{Akmal:1998cf}, SLy \cite{Douchin:2001sv}, LS220 \cite{Lattimer:1991nc}, Shen \cite{Shen:1998gq}, PS \cite{Pandharipande:1975zev}
and PCL2 \cite{Prakash:1995uw}, and one polytropic EoS $\rho=K p^{1/2}$. For QS, three EoS are applied: SQM1, SQM2, and SQM3 \cite{Prakash:1995uw}.
{\renewcommand{\arraystretch}{1.2}
\begin{table}
\begin{centering}
\begin{tabular}{cccccccc}
\hline
\noalign{\smallskip}
 $y_i$ & $x_i$ &&  \multicolumn{1}{c}{$a_i$} &  \multicolumn{1}{c}{$b_i$}
&  \multicolumn{1}{c}{$c_i$} &  \multicolumn{1}{c}{$d_i$} &  \multicolumn{1}{c}{$e_i$}  \\
\hline
\noalign{\smallskip}
 $\bar{I}$ & $\bar{\lambda}^{\rm{(tid)}}$ && 1.47 & 0.0817  & 0.0149 & $2.87\times 10^{-4}$ & $-3.64\times 10^{-5}$\\
 $\bar{I}$ & $\bar{Q}$ && 1.35  & 0.697 & -0.143  & $9.94\times 10^{-2}$ & $-1.24\times 10^{-2}$\\
 ${\bar{Q}}$ & $\bar{\lambda}^{\rm{(tid)}}$ && 0.194  & 0.0936 & 0.0474  & $-4.21\times 10^{-3}$ & $1.23\times 10^{-4}$\\
\noalign{\smallskip}
\hline
\end{tabular}
\end{centering}
\caption{ (Taken from \cite{Yagi:2013bca}). The fitting values in \eqref{log fit} for the I-Love-Q relations, extracted from 7 NS EoS and 3 QS EoS. }
\end{table}
}

\begin{figure}
    \centering
    \includegraphics[width=0.48\textwidth]{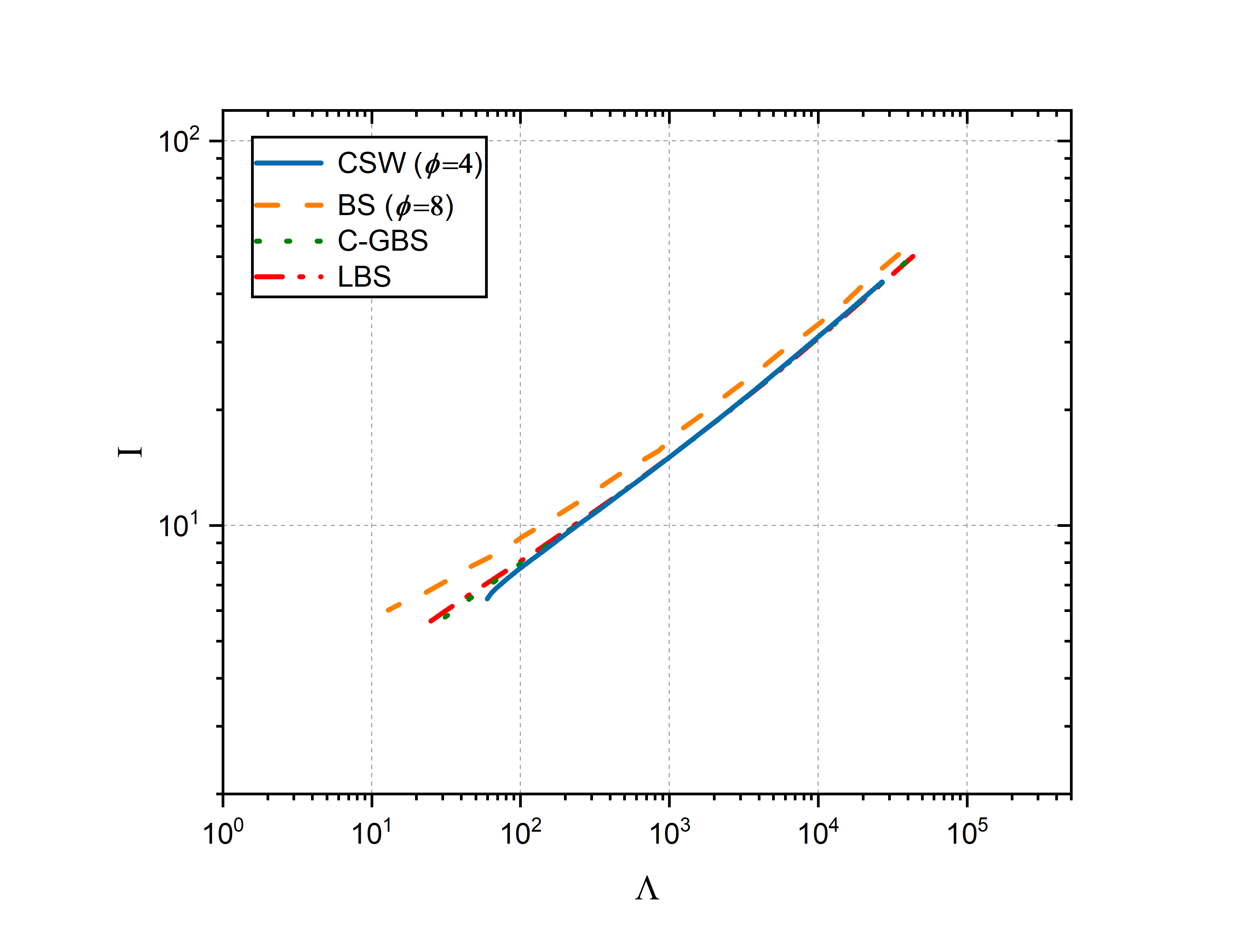}
    \caption{(Taken from \cite{Wu:2023aaz}). The relation between reduced moment-of-inertia $\bar{I}$ and tidal Love number $\bar{\Lambda}$ for 4 DS EoS.}
    \label{fig2}
\end{figure}

We also report that similar relationships exist for DS EoS \cite{Wu:2023aaz}, as the $\bar{I}$-$\bar{\Lambda}$ relation illustrated in Fig. \ref{fig2}, and also the $\bar{I}$-$\bar{Q}$ and $\bar{Q}$-$\bar{\Lambda}$ relations. In fact, for $\phi_n$ EoS, the relation coincides with the neutron star case, with a slight deviation starting from $n=8$. This is predictable because when $p$ is small, they reduce to single polytropic neutron EoS $\rho=p^\gamma$, with $0\leq\gamma \leq 1/2$, and it is the small $p$ part that dominates the behavior of {I-Love-Q} relation.  The Liouville EoS and cosh-Gordon EoS share the same type of {I-Love-Q} relation as the neutron case, implying that the {I-Love-Q} relation is universal for compact stars.

\subsection{Hybrid stars}

Due to the mixture of baryonic matter and DM in the Universe, it is reasonable to speculate the existence of hybrid stars made of both. Since DM almost does not interact with the baryonic ones, and we have little idea of the nature of DM, it is hard to pin down the internal structure of the hybrid stars. By simply classifying the geometric setup, we can have three types of hybrid stars \cite{Zhang:2020pfh, Zhang:2020dfi}. The first type (called Scenario I) is to have a neutron core covered by a DM shell, and the second type (called Scenario II) is to have a DM core covered by a neutron shell. Admittedly, we do not have a mechanism to form a robust domain wall between the core and the shell, such as the one usually adopted by the mechanism of spontaneous symmetry breaking; we assume some unknown mechanism may support such kinds of hybrid stars. Therefore, a more natural type (called Scenario III) is to have mixed baryonic and DM in the inner core but with one of them left in the outer shell. Due to the different inner structures, these hybrid stars should have different mass-radius relations and tidal deformability, which can be distinguished among themselves and from the pure neutron and DS. Of course, this will add up the variety of compact stars and cause more difficulty when identifying the sources of the gravitational events. In \cite{Zhang:2020pfh, Zhang:2020dfi}, all three scenarios have been studied and adopted to fit some GW events. Below, we will review some essential ingredients of hybrid stars in \cite{Zhang:2020pfh, Zhang:2020dfi}.

\subsubsection{Junction conditions}

Because of the domain wall structure inside the hybrid stars of Scenario I and II, we need to impose appropriate junction conditions when solving the TOV configurations and then calculating the associated tidal Love numbers. Before that, the first question is how to determine the inner core's size or the domain wall's position denoted by $r_W$, upon which we impose the junction condition. As we can imagine, the value of $r_W$ should be related to the formation mechanism of the domain wall. Before we have such a mechanism to determine $r_W$ dynamically, we can only treat it as a free parameter in Scenario I and II. On the other hand, in Scenario III, $r_W$ can be determined by solving TOV equations for the pure neutron or DS. 

Given the initial value for the core pressure, we then evolve the single-component TOV equations to obtain the pressure $p_W$ at the domain wall located at $r=r_W$. We require the pressure to be continuous across the domain wall. Then there is a jump of energy density at the domain wall due to the change of the EoS, i.e.,  $\Delta \rho_p=\rho(p_W+0)-\rho(p_W-0)$, or equivalently $\Delta \rho_p=-\left(\rho(r_W+0)-\rho(r_W-0)\right)\equiv -\Delta\rho$ since $p$ decreases as $r$ increases. The discontinuity $\Delta \rho_p$, however, can be determined by requiring the continuity of the sound speed at the domain wall, 
\begin{equation}
	\label{eqdendis}
	\frac{d\rho}{dp}=\frac{1}{c_s^2}=\left. \frac{d\rho}{dp}
	\right|_{p \neq p_W}+\Delta \rho_p \, \delta(p-p_W)\;.
\end{equation}

Similarly, when calculating the TLN for a given TOV configuration of Scenario I and II hybrid stars, we need to integrate \eq{TLN-y} across the domain wall. Only the terms proportional to $\delta$-function can contribute to this integration. Thus, \eq{TLN-y} can be reduced to \citep{Postnikov:2010yn}

\begin{equation}
ry^\prime(r)\big|_{r = r_W}+
r^24\pi e^{\lambda(r)} 
\left({\rho(r)+p(r)}\right)\frac{d\rho}{dp}\big|_{r = r_W}=0\,.
\end{equation}
Making use of $\frac{d\rho}{dp}=\frac{d\rho}{dr}\frac{1}{dp/dr}$ and 
$ \frac{d\rho}{dr}|_{r = r_W}=\Delta\rho \, \delta(r-r_W)$,
we have 
\begin{equation}\label{juntion}
	\Delta y\equiv y(r_W+0)-y(r_W-0)=\frac{\Delta \rho}{p+m(r_W)/(4 \pi r_W^3)}\,.
\end{equation}
Then, the TOV and TLN for scenarios I and II can be solved using the above junction conditions.

\subsubsection{Stability criteria}

After solving a TOV configuration, it is essential to check its stability by studying the linear perturbation. There are various linear modes; the simplest one is radial oscillation, which obeys the master equations derived from the linearized Einstein equation and conservation equation of stress tensor. If there is no growing mode, then the TOV configuration is stable. A detailed derivation for the master equation for a given set of the equation of states of multiple fluids can be found in \cite{Kain:2020zjs}, and its application to DM admixed NS can be found in \cite{Kain:2021hpk}. By the same author, a direct study of the linear stability of the boson stars and hybrid stars based on the scalar-tensor theory can be found in \cite{Kain:2021rmk, Kain:2021bwd}.

If limiting the discussion to a single-fluid star, Here we can introduce a more intuitive but empirical criterion based on the Sturm-Liouville analysis of the radial oscillation eigenmodes to judge the stability of the compact stars based on their mass-radius relation. This is the so-called Bardeen–Thorne–Meltzer (BTM) criteria \cite{1966ApJ...145..505B}. The BTM criteria are stated as follows. In the direction of increasing the core pressure along the mass-radius curve, one stable mode becomes unstable whenever an extremum is passed in the counterclockwise sense. Reversely, an unstable mode becomes stable if an extremum is passed clockwise. The original BTM criteria require transversing the mass-radius curve by starting from stable planet configurations with low enough core pressure. In practice, it is more useful to argue the BTM criteria in a reverse way by traveling along the mass-radius curve to decrease core pressure to avoid the requirement of the above initial condition. We call this Reverse BTM criteria \cite{Zhang:2020dfi, Zhang:2020pfh}, which also applies to the multi-fluid cases, states as follows. Whenever an extremum is passed along the mass-radius curve in either direction of increasing or decreasing the core pressure, a stable mode becomes unstable if the curve bends counterclockwise. Otherwise, an unstable mode becomes stable. 

By applying the Reverse BTM criteria, one can ascertain the unstable regime on the mass-radius curve but can only confirm the stable regime if starting the travel from a stable regime.  Therefore, it is easy to find the stable regime on the $M$-$R$ relation for the pure neutron or DS using the (Reverse) BTM criteria since we know the starting stable region.

\subsubsection{Examples}
Up to the current observations of GW (GW) events, there is insufficient data and accuracy in telling the constituents of an observed compact star candidate. This is because the GW data can give the mass and inaccurate TLN, yielding high degeneracy in the parameter space when fitting the EoS.  Therefore, it will not provide any sensible insight to identify the observed compact star candidate as some neutron, dark or hybrid star. Instead, we should use the GW data to fix the parameter of a given EoS. If we assume the DM model is unique for its associated EoS, the different observed compact candidates should yield the same EoS. 

In \cite{Zhang:2020dfi, Zhang:2020pfh}, we have adopted this strategy to obtain the parameters of some EoS by fitting to some GW events with some possible compact star candidates. For example, in \cite{Zhang:2020dfi}, we have fitted the parameter $\cal B$ of \eq{EoS_4} for the GW190425 based on the Scenario I and II with three different choices of EoSs for the dense neutrons: SLy4 EoS \cite{Douchin:2001sv}, APR4 \cite{Akmal:1998cf} and SKb \cite{Gulminelli:2015csa}. The result is shown in Table \ref{PE}.

\begin{table}[H]
\centering
\begin{tabular}{|c|c|c|c|c|}
\hline
TYPE& \multicolumn{2}{c|}{Scenario I}& \multicolumn{2}{c|}{Scenario II} \\
\hline
& ${\cal B}_4$ & $r_W ( \mbox{km})$&${\cal B}_4 $&$ r_W ( \mbox{km})$\\
\hline
SLy4& $0.07^{+0.09}_{-0.02}$ &$6.44^{+1.98}_{-2.85}$&$0.05^{+0.04}_{-0.02}$&$10.29^{+3.49}_{-4.10}$\\
\hline
APR4& $0.07^{+0.10}_{-0.02}$ &$6.02^{+2.05}_{-2.74}$&$0.05^{+0.04}_{-0.03}$&$9.98^{+3.46}_{-3.69} $\\
\hline
SKb& $0.08^{+0.16}_{-0.03}$ &$6.95^{+2.06}_{-2.89}$&$0.05^{+0.05}_{-0.03}$&$9.78^{+3.65}_{-4.23}$\\
\hline
\end{tabular}
\caption{The inferred best-fitted values of the EoS parameter ${\cal B}_4$ and the core-radius $r_W$ for scenario I \& II of hybrid stars, with three different EoSs for dense neutrons. } \label{PE}
\end{table}

We see that the model parameters ${\cal B}_4$ and $r_W$ are not so sensitive to the chosen EoS for the dense neutrons. Moreover, if we also require the SIDM to explain the dark halo problems, then there is a further constraint on the cross-section $\sigma_{\rm DM}$ of self-scattering, which means \eq{parameter2} should also be satisfied. Combining this constraint with the PE results of Table \ref{PE}, we can put the following constraints on the mass $m$ and the coupling $\lambda_4$ of $\phi^4$ SIDM assuming SLy4 for the EoS of dense neutrons, 
\be
2.68 \mbox{GeV} < m < 10.53 \mbox{GeV}, \qquad 131 < \lambda_4 <3076 \qquad 
\ee
for the scenario I, and 
\be
1.78 \mbox{GeV} < m < 6.65 \mbox{GeV}, \qquad 71< \lambda_4 <1542 \quad 
\ee
for scenario II. The future GW events with the compact star candidates can be used to rule out or reinforce the above prediction.

\subsection{Possible formation mechanism}

\subsubsection{Capture mechanism}
The formation of a DS is an interesting issue, particularly for the potential signals of GW produced by such exotic stellar objects. Here, we provide a particle scale process of DM particles captured by stellar objects. Essentially, this mechanism is difficult for the DS formation since the capture DM number is too small compared with the stellar objects. An estimation for the capture DM mass to be around $10^{11}$~kg for the Sun. The idea is similar to DM direct detection processes. An underground laboratory is needed to search for the signals of the collision between the DM and the detectors' nuclei (or electrons). DM mass and its coupling strength to SM particles are essential in the observation. The stellar DM will scatter with the massive stellar objects (i.e., the Sun or NS, etc.) and be captured by their gravitational wells if the scattered velocity is smaller than the escape velocity. On top of this, for the case that the DM has self-interaction, the capture rate is considerably enhanced~\cite{Chen:2014oaa}. In summary, the number evolution of the captured DM inside the stellar objects is given by 
\begin{align}
\frac{dN_{\chi}}{dt} & =C_{c}-C_{e}N_{\chi}+C_{s}N_{\bar{\chi}}-(C_{a}+C_{se})N_{\chi}N_{\bar{\chi}},\label{eq:n_chi}
\end{align}
depending on various scattering processes among the DM and the stellar objects. Here $C_{c}$ called the capture rate, $C_{e}$
the evaporation rate, $C_{s}$ the capture rate due to self-interaction, $C_{se}$ the self-interaction induced evaporation rate, and $C_{a}$ the annihilation rate. For different underlying assumptions, $C_{c}$ can be divided into spin-independent (SI) interaction and spin-dependent (SD) interaction, respectively, because the distribution of nucleons in the nucleus plays a crucial role. The SD and SI interactions are given by 
\bea
C_{c}^{\rm SD}&\approx&3.35\times10^{24}\Big(\frac{\rho_0}{0.3 \rm GeV/cm^3}\Big) \Big(\frac{270 \rm km/s}{\bar{v}} \Big)^3 \nonumber \\
&&\times \Big(\frac{\rm GeV}{m_{\chi}} \Big)^2 \Big(\frac{\sigma^{\rm SD}}{10^{-6} \rm pb} \Big) \eea
and 
\bea
C_{c}^{\rm SI}&\approx&1.24\times10^{24}\Big(\frac{\rho_0}{0.3 \rm GeV/cm^3}\Big)\Big(\frac{270 \rm km/s}{\bar{v}}\Big)^3 \nonumber \\ 
&&\times \Big(\frac{\rm GeV}{m_{\chi}}\Big)^2 \Big(\frac{2.6\sigma^{\rm SI}_{\rm H}+0.175\sigma^{\rm SI}_{\rm He}}{10^{-6} \rm pb}\Big)    
\eea
respectively, where $\rho_0$ is the local DM density and we take the typical DM density as a reference, $\bar{v}$ is the velocity dispersion if DM is assumed to be thermally equilibrium, $\sigma^{\rm SD(SI)}_{\rm H}$ and $\sigma^{\rm SD(SI)}_{\rm He}$ are the SD(SI) DM-hydrogen and DM-helium scattering cross sections. Because the collisions occur at the non-relativistic limit, the cross-section is inversely proportional to $m^2_{\chi}$. Here $m_{\chi}$ is the DM mass. 


$C_e$ in the second term of Eq.~(\ref{eq:n_chi}) refers to the DM evaporation. This captured DM can be kicked out while scattering with the stellar nucleus. The DM evaporation rate not only depends on the DM mass but also the stellar object's gravitational potential distribution~\cite{Gould:1987ju,Busoni:2013kaa}. We take the Sun as an example; it is expressed as 
\bea
C_e\approx\frac{8}{\pi^3}\sqrt{\frac{2m_{\chi}}{\pi T_{\chi}(\bar{r})}}\frac{v^2_{\rm esc}(0)}{\bar{r}^3}\exp\Big(-\frac{m_{\chi}v^2_{\rm esc}(0)}{2T_{\chi}(\bar{r})}\Big)\Sigma_{\rm evap}, 
\eea 
where $v_{\rm esc}(0)$ is the escape velocity from the core of the Sun, and  $T_{\chi}$ is the DM temperature in the Sun, which corresponds to its average kinetic energy. $\bar{r}$ is the average DM orbit radius, which is the mean DM distance from the solar center, and $\Sigma_{\rm evap}$ is the sum of the scattering cross-sections of all the nuclei within a radius $r_{95\%}$, where the solar temperature has dropped to $95\%$ of the DM temperature. The exponential distribution is due to the DM thermal distribution in the core region of the stellar object. We also take the approximation that the DM temperature is equal to the nucleus temperature around the core. 

$C_s$ is the DM capture rate by colliding off the DM that has been captured inside the stellar objects, and The calculation is similar to the nucleus evaporation effect. One integrates out the final velocity distribution cut by the escape velocity, and it is given by~\cite{Zentner:2009is}
\bea
C_s=\sqrt{\frac{3}{2}}n_{\chi}\sigma_{\chi\chi}v_{\rm esc}(r)\frac{v_{\rm esc}(r)}{\bar{v}}\langle \hat{\phi}_{\chi}\rangle \frac{{\rm erf}(\eta)}{\eta}, 
\eea
where $\langle \hat{\phi}_{\chi} \rangle$ is the dimensionless average of potential for the captured DM inside the stellar object, the value is about $\langle \hat{\phi}_{\chi} \rangle \sim5.1$ for the Sun~\cite{Gould:1991hx}. $n_{\chi}$ is the local density of halo DM, $\sigma_{\chi\chi}$ is the elastic scattering cross section of DM with themselves, $v_{\rm esc}(r)$ is the escape velocity, and $\eta^2$ is the square of the dimensionless velocity of the stellar object in the DM halo. 

Finally, the captured DM in the stellar objects might annihilate each other and produce a pair of SM particles. This effect is described by $C_a$, the annihilation coefficient, given by 
\bea
C_a\approx\frac{\langle \sigma v \rangle V_2}{V^2_1}, 
\eea
where $V_{j}$ is the DM effective volume inside the Sun, it is about $6.5\times10^{28} {\rm cm^3}(\frac{10 \rm GeV}{jm_{\chi}})^{3/2}$  and $\langle \sigma v \rangle$ is the relative velocity averaged annihilation cross section for DM pairs.

\subsubsection{Hydrodynamic approach: Accretion and fragmentation }

One possible formation mechanism of boson stars is the accretion of DM around some massive region due to density fluctuation. The simplest accretion mechanism is Bondi accretion, which assumes spherical symmetry. See Appendix \ref{Bondi_accr} for the outline of solving Bondi accretion. If the DM is nonrelativistic with equation of state $p=\kappa \rho_0^{\gamma}$, then the accretion rate is given by (see also \eq{non_rel_accr_rate})
\be 
\dot{M}=\pi G_N^2 M^2 {\rho_0(\infty)\over c_s^3(\infty)}\Big[{2 \over 5 -3\gamma} \Big]^{(5-3\gamma)\over 2 (\gamma-1)} \qquad
\ee 
where the adiabatic index is restricted to $\gamma\le 5/3$, so that the factor of $\gamma$ is ${\cal O}(1)$ and 
\be 
\dot{M}\sim 10^{-19} M_{\odot} \textrm{yr}^{-1} \Big({M\over M_{\odot}}\Big)^2 \Big({\rho_0(\infty) \over 10^{-21} \textrm{kg}\; \textrm{m}^{-3}}\Big) \Big({c_s(\infty) \over 10^2 \textrm{km}\; s^{-1}}\Big)^{-3}\;.
\ee 
On the other hand, for the relativistic DM, we shall adopt the relativistic Bondi accretion; see Appendix \ref{Bondi_accr} for a brief account. Take the $\lambda |\phi|^4$ self-interacting boson field as an example; the Bondi accretion rate is bounded from below \cite{Feng:2021qkj}, 
\be
\dot{M}_{min}\le \dot{M} < \infty
\ee
with $\dot{M}_{min}$ given in \eq{accr_rate_rel}\cite{Feng:2021qkj}
\bea
&& \dot{M}_{min} := 64 \pi {G_N^2 M^2\over c^3}\rho_B\;,  \nn \\
&\sim & 10^{-9} M_{\odot} \textrm{yr}^{-1} \Big({M\over M_{\odot}}\Big)^2 \Big({\rho_0(\infty) \over 10^{-21} \textrm{kg}\; \textrm{m}^{-3}}\Big) \Big({c_s(\infty) \over 10^2 \textrm{km}\; s^{-1}}\Big)^{-2}\;. 
\eea

From the above result, we can find that the relativistic SIDM enhances the accretion rate by several orders higher than the conventional nonrelativistic WIMP. Despite that, the typical value of $\dot{M}_{min}$ is still smaller than the Eddington accretion rate of baryons $\simeq 10^{-2} M_{\odot} \textrm{yr}^{-1}$.  This implies that it is challenging to form dark boson stars by accreting the DM for either WIMP or SIDM.

If we instead consider the above Bondi accretion of DM around a supermassive BH with mass $\simeq 10^6 \sim 10^9 M_{\odot}$, then the accretion rate will be enhanced by about $10^{12} \sim 10^{18}$ factor. This means that the DM can accumulate quickly around a supermassive BH and form a spike profile. The discussion of the detailed profiles of the spikes can be found in \cite{Feng:2021qkj}. The typical density profile of the spike goes as $r^{-\alpha}$ with $\alpha \simeq 1 \sim 2$ depending on the equation of state and the location around the BH. This sharp spike can easily fragment further due to the usual Jean's instability, and the resultant fragments will be the (seed of) boson stars. This accretion and fragmentation mechanism can speed up the production of DS. The mass distribution for the DS produced from the above mechanism is an interesting issue for future study. 

Finally, we should emphasize that we assume the DM does not interact with photons or some dark photons. This is quite different from the usual Eddington accretion, for which there will be a radiative energy outflow carried by photons. Its typical accretion rate around a BH of mass $M$ is about $10^{-8} M_{\odot} {\rm yr}^{-1} ({M \over M_{\odot}})$ \footnote{See https://ilyamandel.github.io/BackOfTheEnvelopeNotes\\/Eddington.pdf}.

\subsubsection{A tentative example of composite dark matter: Mirror copy approach}
From the above discussions, it is challenging to form DS by accretion mechanism within a reasonable timescale mainly due to the weak interactions with the standard model particles or the constraints on their self-interactions with the observed astrophysical phenomena. 

On the other hand, baryonic stars are easier to form through the interplay of large-scale structure and molecular dynamics. An interstellar cloud of molecules can collapse to form the seed for star formation \cite{schulz2012formation,2011isf..book.....W}.  It is tempting to speculate the possible mechanism for ample DS formations by mimicking the one for the baryonic stars. The simplest way is to speculate the DM model as the mirror copy of the baryonic one, a kind of asymmetric composite DM discussed earlier. For example, we can consider the dark quark models with $SU(3)$ color group so that dark nucleons can form by the confinement mechanism. Suppose we do not mirror the electroweak sector for the DM. In that case, the color-neutral dark nucleons with the mass scale from the dimensional transmutation \cite{Coleman:1973jx} below the GeV scale will be the fundamental constituents from which suitable dark molecules can form by aggregation via the residual van der Waals forces. These dark molecules will be the primary materials to form the interstellar dark clouds for star formations.

Because of lacking the atomic structure in the absence of dark electroweak sectors, the size of the dark nucleon will be five orders smaller than the one of the baryonic atom. Based on naive scaling, this implies that the DS could be at least five orders smaller than the baryonic stars if DS formations follow the same mechanism of molecular dynamics as the baryonic ones. This means that the typical DS with about one solar mass will have compactness like the NS. The DS formed in this way will have a similar internal structure to the NS, except they are composed of dark neutrons. As the underlying formation mechanism is similar to the baryonic ones, we expect the formation rate to be comparable to or even faster than the baryonic ones due to the smaller overall size scale, thus, stronger van der Waals forces. Therefore, we will expect to have populated binary DS merger events to be observed by LIGO/Virgo/KAGRA. Due to their compactness, thus, small tidal deformability, these kinds of DS will behave like mimickers of BH and NS. For future GW detectors with the high capability of multimessenger detection rate, we can tell these DS from the NS by the observed multimessenger signals. On the other hand, these DS can also be adopted to explain the BH mass gap events. Detailed studies of the dark molecule formation and the subsequent cloud formation are needed for a full-scale understanding.

\section{Mimickers of black holes and neutron stars}\label{mimickers_NB}

In gravitational-wave (GW) and electromagnetic (EM) observations, BH are generally justified by their masses and invisibility rather than the determination of zero TLN. In the EM case, getting the TLN information is impossible, while in the GW case, the TLN corrections are usually too weak for distant events. Thus, it is difficult to distinguish DS from BH with similar masses. On the other hand, NS can have detectable TLN but are constrained to have masses between $1 M_{\odot}$
and $2.6 M_{\odot}$. {The lower mass bound is suggested by the optical observations that almost no NS with masses below $1 M_{|odot}$ have been observed \cite{Lattimer:2014wci}. The upper mass bound depends on the EoS of dense nuclear matter, which is uncertain and should be inferred from the observed M-R relation. However, most candidate EOS suggest that the upper mass bound shall not be larger than $2.6 M_{\odot}$ \cite{LIGOScientific:2020zkf,Lattimer:2021emm}.} Therefore, if the boson stars can have masses and TLNs fitted to the above consideration, it is hard to distinguish them from BH or NS.  Such kinds of boson stars can be the mimickers of BH or NS. The masses of the boson stars can also fall in the ranges of either lower or higher mass gaps for the BH, and then these boson stars can be the candidates for the gap events with small TLNs.

We illustrate that BS is the BH and NS mimickers rather than fermionic stars. It is because both EOCs could have similar (or even identical) macroscopic M-R relations, and currently, the only GW observable to differentiate ECO from the BH is the tidal deformability. With the degeneracy of various DM models, one essentially can not pin down the underlying nature of DM, namely fermionic particles or bosonic ones, from a single event. However, suppose we believe a single and unique DM particle exists to explain the cosmological anomaly, statistical-wise. In that case, one can read out the details of the underlying DM model after accumulating sufficient events and data.

In this section, we would like to demonstrate the possibility of the boson stars as the mimickers of BH or NS based on the EoSs discussed in section \ref{EoS_sec}.  

\subsection{Mimickers in the lower mass range}

\begin{figure}[H]
    \centering
    \includegraphics[width=0.46\textwidth]{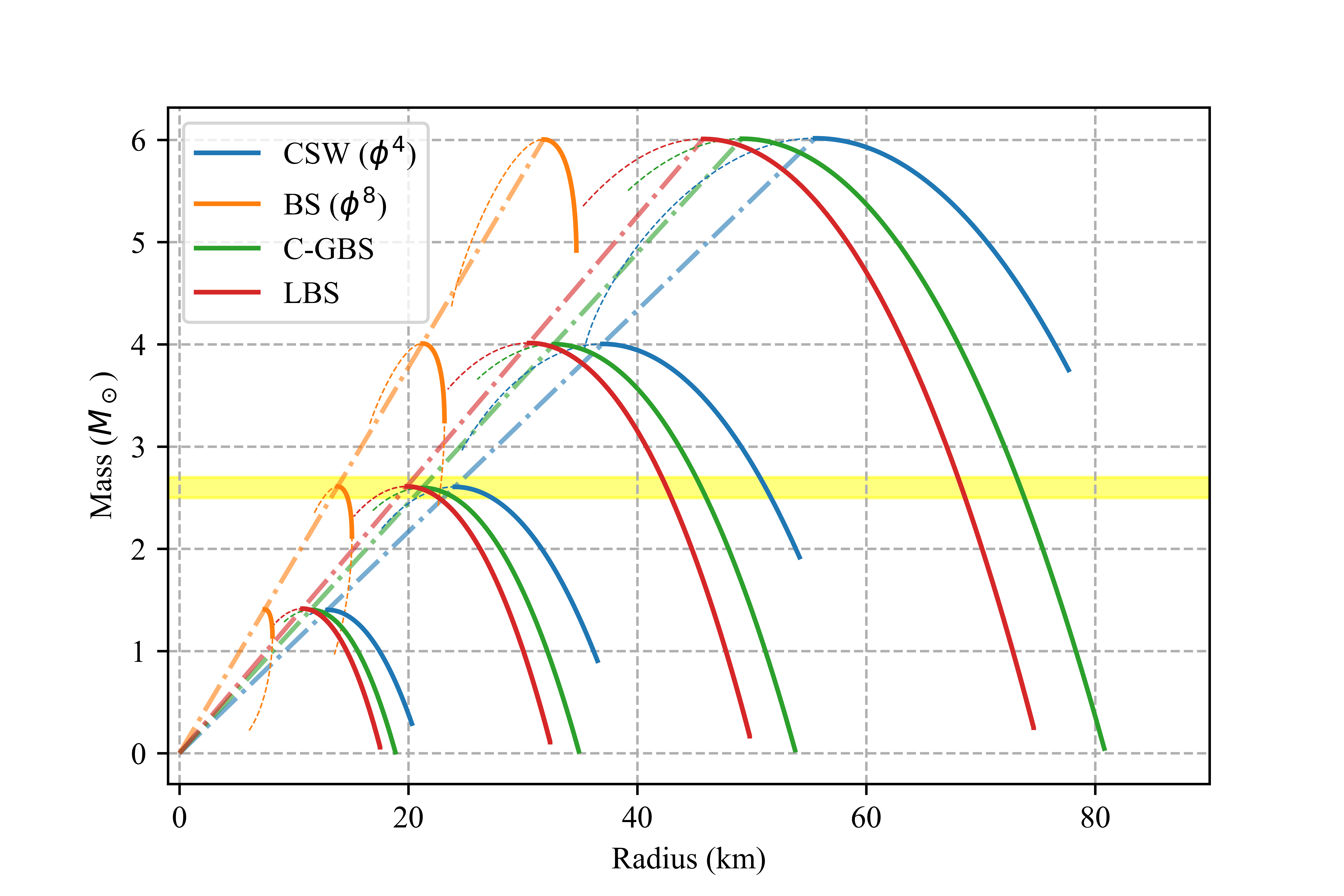}
    \caption{Mass-Radius relation for the four types of the considered boson stars in the lower mass gap region. Blue curves stand for the General BS with $n = 4$ (CSW $\phi^4$), orange curves with $n = 8$ (BS $\phi^8$), green for the Cosh-Gordon BS (C-G BS), and red for the Liouville BS (LBS). These groups of solid curves are marked, with maximum masses to be $1.4M_\odot,\ 2.6M_\odot,\ 4M_\odot,$ and $6M_\odot$, respectively. The dash-dot line with the same color as the corresponding EoS links the maximal-mass points of the same type of boson stars labeled by the same color, including those not drawn explicitly. The yellow area is the range of GW190814 secondary mass($m_2$) $2.6^{+0.1}_{-0.1}M_\odot$ at $90\%$ {credible interval from GW Open Science Center (GWOSC) \cite{TheLIGOScientific:2017qsa}.} The dashed curve region of each curve is unstable. Different M-R curves are obtained by tuning the parameter $\cal B$ of the corresponding EoS, shown in Table \ref{tab:B_parameter}.}
    \label{fig:mass_radius_lower}
\end{figure}

\begin{figure}[H]
    \centering
    \includegraphics[width=0.46\textwidth]{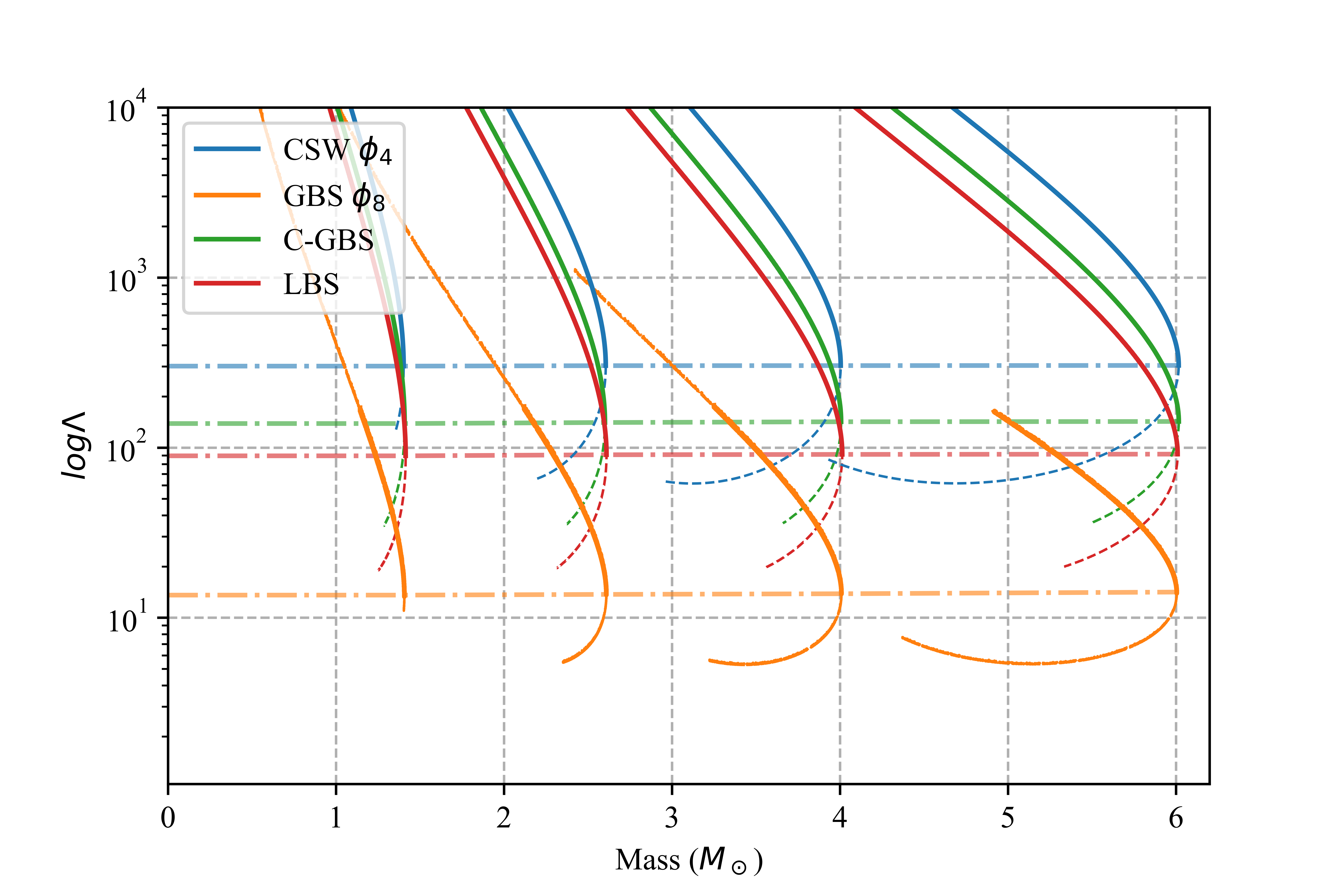}
    \caption{TLN-Mass relation for the four types of the considered boson stars in the lower mass gap region. The color and mark schemes are the same as in Fig. \ref{fig:mass_radius_lower}.
    Different TLN-M curves are obtained by tuning the parameter $\cal B$ of the corresponding EoS, shown in Table \ref{tab:B_parameter}.}
    \label{fig:L_mass_lower}
\end{figure}

We start with the cases for the lower mass gap.  Fig.  \ref{fig:mass_radius_lower} and Fig. \ref{fig:L_mass_lower} show the Mass-Radius and TLN-Mass relations, respectively, with the masses below $6 M_{\odot}$ for the EoSs discussed in section \ref{EoS_sec}, namely, $\phi^4$ (CSW) BS, $\phi^{8}$ BS, Liouville BS, and Cosh-Gordon BS. By varying $\mathcal{B}$ or $\mathcal{B}_n$ in the EoS, see the range shown in Table \ref{tab:B_parameter}, the stable boson stars can cover a wide mass range.

The $2.6\ M_\odot$ curves are marked to compare with GW190814, whose secondary star has a mass $2.59^{+0.08}_{-0.09}\ M_\odot$, with no measurable TLN data, and no EM counterpart of this event is reported. The General BS EoS $\phi^n$ could serve as the candidates for this component compact object. Similarly, due to the wide parameter space of boson star models, now LBS, C-G BS also could achieve $2.59\ M_\odot$ while keeping a low TLN as shown in Fig. \ref{fig:L_mass_lower}.

The $4 M_{\odot}$ curves show the typical candidates for the lower gap events, with the radius ranging from about $22$ km to $36$ km and the TLN from about $10$ to $250$. Since masses of the gap events exceed the possible upper mass range of the NS while still too low for generally accepted BH scenarios, boson stars are the very competitive candidates. Several other compact objects are observed in the lower gap, including GW190814.  One of them is named 2MASS J05215658+4359220 \cite{Thompson:2018ycv}, where a  red giant forms a non-interacting binary system with a likely BH,  which has a mass of $3.3^{+2.8}_{-0.7}$ $M_{\odot}$.

\begin{table}[H]
    \centering
    \begin{tabular}{lllll}
    \hline\hline
    \multirow{2}{5.5em}{Parameter}  &  \multicolumn{4}{c}{$M_{\mathrm{max}}$}\\
    & $1.4M_\odot$ & $2.6M_\odot$ & $4.0M_\odot$ & $6.0M_\odot$ \\
    \hline 
    CSW ($\mathcal{B}_\mathrm{4}$) & $0.086$ & $0.047$ & $0.031$ & $0.02$ \\
    BS $\phi^8$ ($\mathcal{B}_\mathrm{8}$) & $0.10$ & $0.055$ & $0.036$ & $0.024$ \\
    LBS ($10^4 \mathcal{B}_\mathrm{L}$) & $27.5$ & $8.05$ & $3.40$ & $1.52$ \\
    C-GBS ($10^5 \mathcal{B}_\mathrm{C}$) & $40.0$ & $11.2$ & $4.90$ & $2.18$ \\
    \hline\hline
    \end{tabular}
    \caption{The values of $\mathcal{B}$ of the corresponding EoS for the curves shown in Fig. \ref{fig:mass_radius_lower} and Fig. \ref{fig:L_mass_lower} below.}
    \label{tab:B_parameter}
\end{table}

From the results, we see that all four boson star models can serve as the mimickers of NS for the masses between $1 M_{\odot}$ and $2.6 M_{\odot}$ as their TLNs for stable configurations are below few hundred, that is comparable with the current observation from GW170817 \cite{TheLIGOScientific:2017qsa, Abbott:2018wiz}. By the multimessenger constraints \citep{Dietrich:2020efo}, a neutron star of 1.4 $M_{\odot}$  should have a radius $11.75^{+0.86}_{-0.81} \mbox{ km}$ at $90\%$ confidence level. Thus, we find that the LBS and C-G BS curves with maximum 1.4 $M_{\odot}$ fall into this range, while $\phi^4$ BS, $\phi^{8}$ BS seem outside of it. But since the mass does not necessarily be the maximum, $\phi^{8}$ BS  curve with higher maximum mass, say 2 $M_{\odot}$, could clear this criterion.

The above Mass-Radius relations can also explain the secondary mass $2.59^{+0.08}_{-0.09}$ $M_{\odot}$ of GW190814 as the mimicker of a BH in the lower mass gap. Therefore, by choosing the EoS parameter $\mathcal{B}$ or $\mathcal{B}_n$, the boson star models can easily accommodate the events of black-hole and neutron-star mimickers, even for the ones in the lower mass gap. This comes as no surprise as there are almost no constraints on the properties of DM, hence boson stars. Instead, one can use the GW events, especially the ones in the mass gap, to constrain the properties of DM, such as the SIDM considered here.

Besides, two more remarks are in order. First, the boson star models can also predict compact stars with masses less than $1 M_{\odot}$. For example, one $0.6 M_{\odot}$ would have a radius from $3$km for $\phi^8$ BS to $5.5$km for $\phi^4$ BS. These will be novel candidates for GW detections. {See \cite{Nitz:2022ltl, LIGOScientific:2022hai} for more discussions on the search plan.} Second, the maximal compactness is a constant for each EoS, independent from the adjusting parameter $\mathcal{B}_n$ or $\mathcal{B}$. As shown in  Fig. \ref{fig:mass_radius_lower}, if we link the maximal-mass points of MR curves with the same EoS together by dashed lines, they form strict straight lines, passing through the original point. This can be explained by the scaling symmetry of the TOV equation as discussed in section \ref{scaling_sec}.

\subsection{Mimickers in the higher mass range}

\begin{figure}[H]
    \centering
    \includegraphics[width=0.46\textwidth]{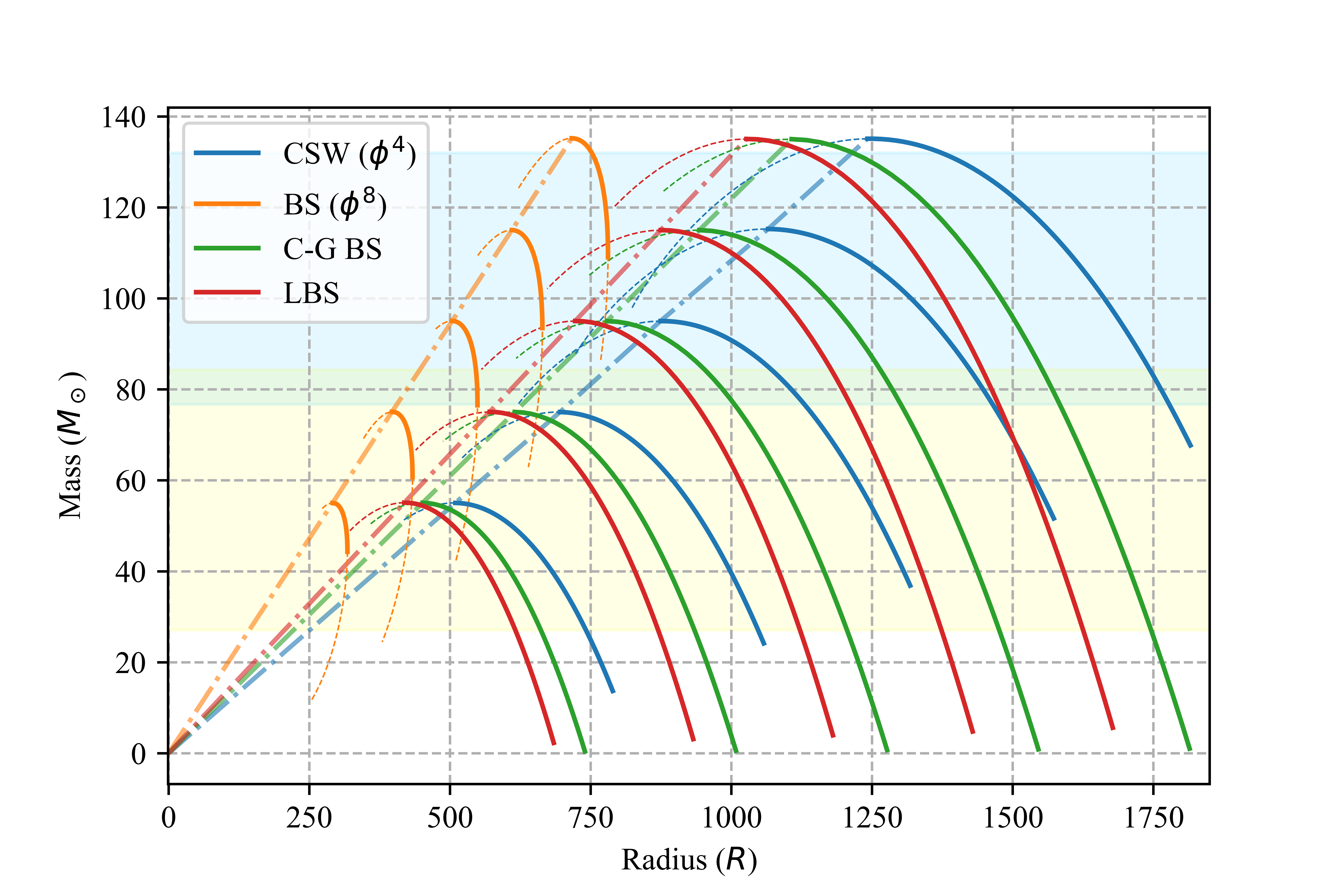}
    \caption{Mass-Radius relation for the four types of the considered boson stars in the upper mass gap region. 
    These groups of solid curves are marked, with maximum masses of $55 M_\odot,\ 75M_\odot,\ 95M_\odot, 115M_\odot$, and $135M_\odot$, respectively. The dash-dot line links the TLNs ($\Lambda$s) of the maximal-mass points of the same type of boson stars labeled by the same color, including those not drawn explicitly. The sky blue area is the range of GW190521 primary mass($m_1$) $98.4^{+33.6}_{-21.7} M_\odot$. The yellow area is the range of GW190521 secondary mass($m_2$) $57.2^{+27.1}_{-30.1} M_\odot$. Both data are at a $90\%$ confidence level. }
    \label{fig:mass_radius_upper}
\end{figure}

\begin{figure}[H]
    \centering
    \includegraphics[width=0.46\textwidth]{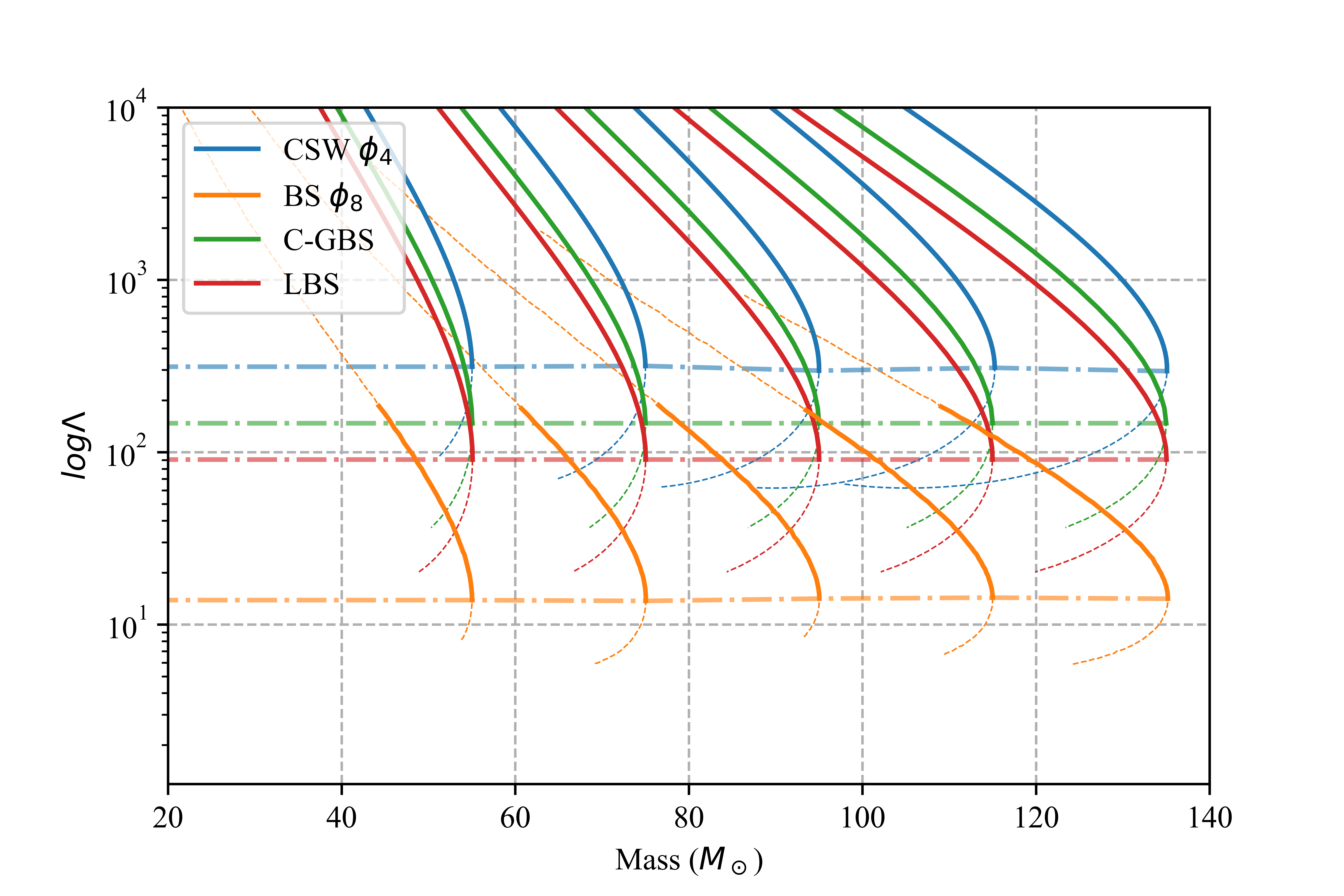}
    \caption{TLN-Mass relation for the four types of the considered boson stars in the upper mass gap region. The color and mark schemes are the same as in Fig. \ref{fig:mass_radius_upper}.   
    Different TLN-M curves are obtained by tuning the parameter $\cal B$ of the corresponding EoS, shown in Table \ref{tab:B_parameter_upper}.}
    \label{fig:L_mass_upper}
\end{figure}

We now consider the cases of black-hole mimickers around the upper mass gap, i.e., $50$ to $150 M_{\odot}$. We show the Mass-Radius and TLN-Mass relations in Fig. \ref{fig:mass_radius_upper} and Fig. \ref{fig:L_mass_upper}, respectively, in the range of the upper mass gap. The corresponding range of the EoS parameter $\cal B$ is also shown in Table \ref{tab:B_parameter_upper}. 

As expected, we find that the boson stars can be the black-hole mimickers in the upper mass gap by choosing the proper EoS parameter $\cal B$. Thus, they can be the candidates to explain the GW events in the upper mass gap as the black-hole mimicker,  { such as the primary mass of GW190521, which is estimated to be $85.3^{+21}_{-14}$ $M_{\odot}$ with the final mass of merger to be  $142^{+28}_{-16}$ $M_{\odot}$ \cite{190521}. }
All four EoS could reach this primary mass, with TLN as low as 100 to 300. This makes them nice black-hole mimickers because they are indistinguishable from BH, considering their low TLN and little interaction with EM signals.  {There are alternative proposals for explaining GW190521 as the BH mimickers based on different boson star models, e.g., in \cite{CalderonBustillo:2020fyi}, it is interpreted as the head-on collision of vector boson 
(Proca) stars; see also \cite{CalderonBustillo:2022cja}.}

\begin{table}[H]
    \centering
    \begin{tabular}{clllll}
    \hline\hline
    \multirow{2}{1.5em}{$\cal B$}  &  \multicolumn{4}{c}{$M_{\mathrm{max}}$}\\
    & $55 M_\odot$ & $75 M_\odot$ & $95 M_\odot$ & $115 M_\odot$ & $135 M_\odot$\\
    \hline
    $10^3 {\mathcal B}_{\mathrm 4}$ & $2.23 $ & $1.64 $ & $1.30$ & $1.07$ & $0.908$ \\
    $10^3 {\mathcal B}_{\mathrm 8}$ & $2.61$ & $1.91$ & $1.51$ & $1.25$ & $1.06$ \\
    $10^8 {\mathcal B}_{\mathrm L}$ & $18.0$ & $9.70$ & $6.05$ & $4.13$ & $3.00$\\
    $10^9 {\mathcal B}_{\mathrm C}$& $26.0$ & $14.0$ & $8.73$ & $5.96$ & $4.32$\\
    \hline\hline
    \end{tabular}
    \caption{The values of $\mathcal{B}$ of the corresponding EoS for the curves shown in Fig. \ref{fig:mass_radius_upper} and Fig. \ref{fig:L_mass_upper}}
    \label{tab:B_parameter_upper}
\end{table}

Besides, many other events contain stars in the upper mass gap, like GW190403$\_$051519 with an $88.0^{+28.2}_{-32.9}$ $M_{\odot}$ primary star, GW190426$\_$190642 with a $106.9^{+41.6}_{-25.2}$ $M_{\odot}$ primary star and a $76.6^{+26.2}_{-33.6}$ $M_{\odot}$ secondary star, GW190929$\_$012149 with an $80.8^{+33.0}_{-33.2}$ $M_{\odot}$ primary star, and GW200220$\_$061928
with an $86^{+38}_{-22}$ $M_{\odot}$ primary star, etc.  {It seems that upper-mass-gap GW events are not rare. However, we shall emphasize that none of the above events have enough significance to be included in the LVK's testing GR analysis. Despite that, we may expect similar mass-gap events to occur in the next-generation GW detector.} Of course, these mass-gap events could be either primary or secondary objects by merging primary compact objects. However, the latter possibility should be far less than the former one,  {because most mergers do not have their final masses in the upper mass gap. Moreover, there are ways to distinguish the secondary from the primary objects, notably by the larger spin of the secondary from the orbital angular momentum before the merger.}

Also, note that using the same EoS with fixed parameter $\cal B$ cannot explain the lower and upper mass gap events simultaneously. To do the job, at least two different SIDMs with quite different $\cal B$. Despite that, the SIDM is still the most simple and natural to yield the black-hole mimickers for the gap events. After introducing the data analysis methodology in the next section, we will present some posteriors of GW190521 and GW190814 to fit the models of BH mimickers.

\section{Methodology of data analysis}\label{data_analysis}

Due to unavoidable noises in the GW detector, it is essential to devise the data analysis methodology to extract the source properties from the observed strain data. Also, the GW analysis heavily relies on the waveform models. If one would like to consider the GW signals from the DS, the construction of the waveforms will be dictated by the underlying DM theory. Therefore, it is impossible to give a full review on this subject. Instead, we will sketch the basic concepts of the data analysis methodology and how to discriminate the BH and NS from their mimickers. We hope this will be helpful for the novices. For the recent progress, please see \cite{Maselli_2017, Sennett:2017etc, Zhang:2020pfh, Zhang:2020dfi, Vaglio:2023lrd, Narikawa:2021pak} and the references therein and their future citations.

Below, we first review the general theory of Bayesian statistics  \cite{gelman2013bayesian, statrethinkingbook, kruschke2015doing} on which the framework of GW data analysis follows. Denote the observed data as a vector, $y=(y_1, y_2,\ldots,y_n)$ with the index labeling the various observed properties. Similarly, 
the parameters of interest 
can be written as $\bm{\vartheta} = (\vartheta_1, \vartheta_2,\ldots,\vartheta_n)$. 
Due to the detector's noises, the measured $\vartheta$ is a random variable. For example, $\vartheta\sim N(\mu, \sigma^2)$ or $p(\vartheta) = N(\vartheta|\mu, \sigma^2)$, denotes that $\vartheta$ obeys a (multi-)normal distribution with mean $\mu$ and variance $\sigma^2$ which is represented by $N(\vartheta|\mu, \sigma^2)$. Here are two key elements of Bayesian statistics. 

\bigskip

\paragraph{Bayes' theorem---} The joint probability for $\vartheta$ and $y$ can be written as 
\begin{align}\label{Eq:joint probability}
p(\vartheta, y) = p(\vartheta)p(y|\vartheta),
\end{align}
where we call $p(\vartheta)$ and $p(y|\vartheta)$ the prior and sampling distributions respectively.
Bayes' theorem is a basic property of conditional probability on the known data $y$, which can be given by 
\begin{align}\label{Eq:Bayes' theorem}
p(\vartheta|y) = \frac{p(\vartheta, y)}{p(y)},
\end{align}
where 
\begin{align}
p(y) = \sum_{\vartheta}p(y|\vartheta)
\quad\text{or}\quad  
p(y) = \int d\vartheta\, p(y|\vartheta)
\end{align}
is called the marginalized distribution of data or evidence.
By substituting \eqref{Eq:joint probability} into \eqref{Eq:Bayes' theorem}, we obtain the so-called posterior distribution
\begin{align}\label{Bayes_1}
p(\vartheta|y) = \frac{p(\vartheta)p(y|\vartheta)}{p(y)}\, \propto \, p(\vartheta)p(y|\vartheta).
\end{align}
This tells the probability distribution of the parameter $\vartheta$ extracted from the data $y$. 

Based on \eq{Bayes_1} with a given prior $p(\vartheta)$, to infer the posterior $p(\vartheta|y)$ of the source properties $\vartheta$ for a given data $y$, we need to construct a probability model from $y$ to obtain the likelihood function $p(y|\vartheta)$. The standard method to construct the likelihood function is by the numerical methods based on Monte-Carlo methods, such as Markov chain Monte-Carlo (MCMC) or nested sampling. We will not discuss these methods here.

The posterior predictive distribution can be identified as the average over the posterior distribution of $\vartheta$, the distribution $p(\vartheta|y)$
\begin{align}
p(\tilde{y}|y) = \int d\vartheta\, p(\tilde{y}|\vartheta) p(\vartheta,y),
\end{align}
where we have
\begin{align}
p(\tilde{y}|\vartheta) = p(\tilde{y}|\vartheta, y) 
\end{align}
due to the independence of $\tilde{y}$ and $y$.

\bigskip

\paragraph{Bayes factor---} Scientific measurements aim to verify or falsify some scientific theory, scenario/hypothesis. Usually, there exist competitive theories/hypotheses. To deal with the so-called hypothesis testing, we can evaluate the Bayes factor to see which theory/hypothesis fits the data better. Let us denote two competing hypotheses as $H_A$ and $H_B$. For each model hypothesis $m=A,B$, we have the corresponding prior $p(\theta_m|H_m)$ and likelihood $p(y|\theta_m,H_m)$, then the associated posterior by Bayes' theorem should be 
\begin{align}
p(\vartheta_m|y,H_m)\, \propto\, p(y|\vartheta_m,H_m)p(\vartheta_m|H_m).
\end{align}

On the other hand, if we like to compare the superiority between the hypotheses $H_{A, B}$ based on the observed data $y$, we would like to evaluate the following ratio of odds,
\begin{align}
\frac{p(H_A | y)}{p(H_B | y)}
= \frac{p(H_A)}{p(H_B)}
\frac{p(y | H_A)}{p(y | H_B)},
\end{align}
where we have again adopted Bayes's theorem. The part that involved both data and hypothesis is called the Bayes factor,
\begin{align}\label{Eq:bayes factor}
\mathcal{B}_{AB}
= \frac{p(y | H_A)}{p(y | H_B)}
= \frac{\int d\vartheta_A\, p(\vartheta_A | H_A) p(y | \vartheta_A,H_A)}{
  \int d\vartheta_B\, p(\vartheta_B | H_B) p(y | \vartheta_B,H_B)}\;. 
\end{align}
If we do not prefer either hypothesis $H_A$ or $H_B$, we can assume ${p(H_A) = p(H_B)}$. Thus, the ratio of odds is nothing but the Bayes factor. From the second equality of \eq{Eq:bayes factor}, we see that it needs to obtain the likelihood function for each model hypothesis to evaluate the Bayes factor, which can be done by invoking either MCMC or nested sampling.

In the above, we have reviewed the basics of Bayesian statistics. In the following, we will apply GW data analysis.  
   
\subsection{Matched filtering}
The data analysis aims to extract the source properties, such as the masses, spins, and locations of the binary compact objects, from the observed strain data. Also, we need to prepare some mock strain data for the overhead training of the data analysis algorithm.  {Now we denote the time series strain data by $\bm{d}(t) = \{d_i\}$, which uniformly sammpled at times $t_i$, instead of $y$ in the GW data analysis.} In general, a strain data $\bm{d}(t)$  can be decomposed as following, 
\begin{align}
    \bm{d}(t) = \bm{n}(t) + \bm{h}_{\bm{\vartheta}}(t).
\end{align}
where $\bm{n}(t)$ is the intrinsic detector's noise, which contaminates the intrinsic/theoretical strain $\bm{h}_{\bm{\vartheta}}(t)$ from the sources in the strain/mock data. We usually assume $\bm{n}(t)$ is a stationary Gaussian noise, though the deviation from this assumption is expected but shall be controllable. With such an assumption, one should fit the best fit of $\bm{d}(t)- \bm{h}_{\bm{\vartheta}}(t)$ to a Gaussian by scanning the space of the theoretical strain $ \bm{h}(t)$ to construct the likelihood function $p(\bm{d}(t)|\bm{\vartheta}, H)$ for the source parameters $\bm{\vartheta}$ in the model $H$. This is the so-called matched filtering method, which aims to maximize the  signal-to-noise ratio (SNR) given by
\be 
\sup_{\bm{\vartheta}, t_0} \Re \int_{f_{\rm min}}^{f_{\rm max}}{\tilde{\bm{d}}(f) \cdot \tilde{\bm{h}}_{\bm{\vartheta}}^*(f) \over S_n(f)} e^{2\pi i f t_0} df
\ee
where the tilde quantity denotes its Fourier transform counterpart, and $S_n(f)$ is the power spectral density of the detector noise.  

\subsection{Bayesian inferences for gravitational wave data analysis}
 To avoid discussing the detection problem, the matched filtering method can be adopted for the low-latency search for the GW event candidates by setting an SNR threshold, say SNR $\ge 10$. Once a GW event candidate is found, the next step is to perform parameter estimation for the source properties by scanning the likelihood function based on MCMC or nested sampling with the help of matched filtering.  

The parameter estimation can help us understand the parameter space's credible region. In GW astronomy, there are many open-source toolkits for parameter estimation, e.g.,  {LALInference\cite{swiglal}, PyCBC Inference \cite{Biwer:2018osg},} and BILBY\cite{Ashton:2018jfp}. In this paper, we choose the PyCBC as the main toolkit. 
The PyCBC is based on a Python code environment for parameter estimation for compact binary coalescence (CBC) signals; it uses Bayesian inference to infer the properties of the source.  {Given the observed data $\bm{d}(t)$ under the hypothesis $H$, 
the Bayes's theorem \eqref{Bayes_1} can be read as}
\begin{align}
    p(\bm{\vartheta}|\bm{d}(t), H) = \frac{p(\bm{d}(t)|\bm{\vartheta}, H) p(\bm{\vartheta}|H)}{p(\bm{d}(t)|H)}
    \label{Bayes_theorem}
\end{align}
where $p(\bm{\vartheta}|\bm{d}(t), H)$ is the posterior probability density describing the conditional probability that the signal has parameter $\bm{\vartheta}$ given observation data $\bm{d}(t)$ and  {an uncertain hypothesis $H$, for example, it includes waveform model, Gaussian noise, etc. The probability density} $p(\bm{d}(t)|\bm{\vartheta}, H)$ is the likelihood function, which is the probability of obtaining the observation data $\bm{d}(t)$ given the  {hypothesis} $H$ with parameter $\bm{\vartheta}$. $p(\bm{\vartheta}|H)$ is prior describing our knowledge about the parameter $\bm{\vartheta}$ before considering the observation data $\bm{d}(t)$.

\par If we are interested in a parameter $\bm{\vartheta}$, we can marginalize the posterior probability by integrating $p(\bm{d}(t)|\bm{\vartheta}, H)p(\bm{\vartheta}|H)$ over the unwanted parameters. If we marginalize over all parameters, we can get the evidence $p(\bm{d}(t)|H)$
\begin{align}
    \int^{\vartheta_1^\mathrm{max}}_{\vartheta_1^\mathrm{min}} \ldots \int^{\vartheta_N^\mathrm{max}}_{\vartheta_N^\mathrm{min}} d\vartheta_1 \ldots d\vartheta_N\ & p(\bm{d}(t)|\bm{\vartheta}_i, H)p(\bm{\vartheta}_i|H) \notag \\
    &= p(\bm{d}(t)|H)
\end{align}
The evidence is a normalizing constant for the posterior probability given model $H$. If we want to compare two different models, $H_A$ and $H_B$, the Bayes factor can be used to determine which model favored  {according to the Bayes factor \eqref{Eq:bayes factor}.}
If  {$\mathcal{B}_{AB}$} is greater than $1$, then the model $H_A$ is favored over $H_B$.

\par To perform Bayesian inference, we implement the MCMC algorithm { \cite{metropolis;1953, geman;1984, gilks;1995}} to sample the parameter space and then calculate the posterior probability density.  
After the first $l$ initial iterations, the $k$-th Markov chain yields the set of parameter{s}, $\bm{\vartheta}^{(k)}_l$ given the prior probability density function. The sampling algorithm will yield a new proposed set of parameters, $\bm{\vartheta}^{(k)}_{l'}$ with probability $P(\bm{\vartheta}^{(k)}_{l}, \bm{\vartheta}^{(k)}_{l'})$. Suppose a new set of parameters is chosen. In that case, the sampler then computes an acceptance probability, $\gamma$, which determines if the Markov chain should move to the proposed parameter set $\bm{\vartheta}^{(k)}_{l'}$, such that $\bm{\vartheta}^{(k)}_{l+1} = \bm{\vartheta}^{(k)}_{l'}$. Otherwise, $\bm{\vartheta}^{(k)}_{l'}$ will be rejected, then $\bm{\vartheta}^{(k)}_{l+1} = \bm{\vartheta}^{(k)}_l$. After sufficient iterations, the ensemble converges to a distribution proportional to the posterior probability density function sampling. The true astrophysical parameters can be estimated from a histogram of the Markov chain distribution of the samplers in the parameter space.

\par {  In PyCBC, there are several well-developed software packages that implement algorithms for sampling from posterior probability density functions. The commonly used algorithms are \texttt{emcee}, published by Foreman-Mackey \cite{foreman:2013, foreman:2018}, as well as its parallel-tempered version, \texttt{emcee-pt} sampler \cite{vousden:2015}. These sampling algorithms advance the position of the Markov chain based on its previous position and provide inference capabilities in PyCBC when updating these positions.

\par The \texttt{emcee-pt} sampler is a parallel-tempered sampler that improves the exploration of different regions in high-dimensional parameter space by sampling at multiple temperatures. It utilizes a set of Markov chains, where each chain corresponds to a different temperature, $T$, such that 
\begin{align}
    p_T(\vartheta|d(t), H) = \frac{p(d(t)|\vartheta, H)^{\frac{1}{T}}p(\vartheta|H)}{p(d(t)|H)}
\end{align}
These chains exchange states between different temperatures, enabling exploration in the temperature space. The \texttt{emcee} sampler performs the sampling using one temperature, where $T = 1$. 

\par The output from these sampling algorithms is a Markov chain. The consecutive steps of these chains are not independent, as the Markov process depends on the previous states \cite{christensen;2004}. The autocorrelation length $\tau_K$ of the Markov chain is a measure of the number of iterations required to obtain independent samples from the posterior probability density function \cite{madras;1988}. The autocorrelation length of the $k$-th Markov chain, $X^{(k)}_l = \{\vartheta^{(k)}_g; 1 < g < l\}$, of length, $l$, obtained from the sampling algorithm is defined as
\begin{align}
    \tau_K = 1 + 2 \sum^K_{i=1} \hat{R}_i,
\end{align}
where $K$ is the first iteration along the Markov chain, the condition $m\tau \leq K$ is true, $m$ being a parameter which in PyCBC Inference is set to $5$ \cite{madras;1988}. The autocorrelation function, $\hat{R}_i$, is defined as
\begin{align}
    \hat{R}_i = \frac{1}{l\sigma^2}\sum^{l-i}_{t = 1} (X_t - \mu)(X_{t+i}-\mu),
\end{align}
where $X_t$ are the samples of $X^{(k)}_l$ between the $0$-th and the $t$-th iteration, $X_{t+i}$ are the samples of $X^{(k)}_l$ between the 0-th and the $(t+1)$-th iterations. Here, $\mu$ and $\sigma^2$ are the mean and variance of $X_t$, respectively.

\par The initial position of a Markov chain can affect its subsequent positions. The length of time before the Markov chain is considered to have lost any memory of its initial position is referred to as the "burn-in" period. A common practice in MCMC analysis is to discard samples from the burn-in period to prevent any bias introduced by the initial position of the Markov chain on parameter estimation in MCMC. There are several ways in PyCBC Inference to determine when a Markov chain has passed the burn-in period. Among them, we often use the following two methods, \texttt{max\_posterior} and \texttt{n\_acl}.

\par The \texttt{max\_posterior} algorithm is an implementation of the burn-in test used for the MCMC sampler \cite{venitch;2015}. In this method, the $k$-th Markov chain is considered to be past the burn-in period at the first iteration, $l$, for which
\begin{align}
    \log\mathcal{L}^{(k)}_{l} (\vartheta) \leq \max_{k, l}\log\mathcal{L}(\vartheta) - \frac{N_p}{2},
\end{align}
where $\mathcal{L}(\vartheta)$ is the prior-weighted likelihood,
\begin{align}
    \mathcal{L}(\vartheta) = p(d(t)|\vartheta, H)p(\vartheta|H),
\end{align}
and $N_p$ is the number of dimensions in the parameter space. The maximization, $\max_{k, l}\log\mathcal{L}(\vartheta)$, is carried out over all Markov chains and iterations. The ensemble is considered past the burn-in period at the first iteration, where all Markov chains pass this test. 

\par In addition to the \texttt{max\_posterior} test, the \texttt{n\_acl} test is also used. All Markov chains converge when we use the \texttt{max\_posterior} test to return an iteration in MCMC and use it with the \texttt{emcee\_pt} sampler. This underestimates the burn-in period as the initial points still influence the samples from the posterior probability density function. However, adding at least one autocorrelation length to the computation of the burn-in period using the \texttt{max\_posterior} test can reduce this effect. Therefore, we use the \texttt{n\_acl} test in conjunction with the \texttt{emcee\_pt} sampler. If the length of the chains exceeds ten times the autocorrelation length, this test assumes that the sampler has passed the burn-in period. The autocorrelation length is computed using samples from the latter half of the Markov chain. If the test is satisfied, it is considered that the sampler has passed the burn-in period at the midpoint of the Markov chain.
}

\subsection{Some example posteriors for mass gap events as black hole mimickers}

To demonstrate the results of the GW data analysis based on the above Bayesian inference framework, we show the posteriors for two mass gap events,  {GW190521 and GW190814 \cite{LIGOScientific:2023vdi,RICHABBOTT2021100658},} by fitting to the model of BH mimickers, namely, compact stars with nonzero TLN. Here, we use the public data on the GW Open Science Center (https://www.gw-openscience.org) released by the LVC. \footnote{ The version of strain data we used was released on March 9, 2021.}

 To demonstrate the result efficiently, we use the ``IMRPhenomPv2\_NRTidal'' waveform model \cite{Dietrich:2019kaq} instead of the ``SEOBNRv4\_ROM\_NRTidalv2'' model.
Moreover, the BH mimickers have nonzero TLNs, allowing one to distinguish them from BBH.
The analysis of BH mimickers with GW observations shows that the EoS can be constrained by masses and TLNs of a binary \cite{Johnson-Mcdaniel:2018cdu}. In our work, we are also interested in a binary of the BH mimickers, and therefore, we include the analysis of the TLNs in GW190521 and GW190814 without setting both values to zero.

\subsubsection{Upper Mass Gap: GW190521}

\begin{figure}[H]
    \centering
    \includegraphics[width=0.48\textwidth]{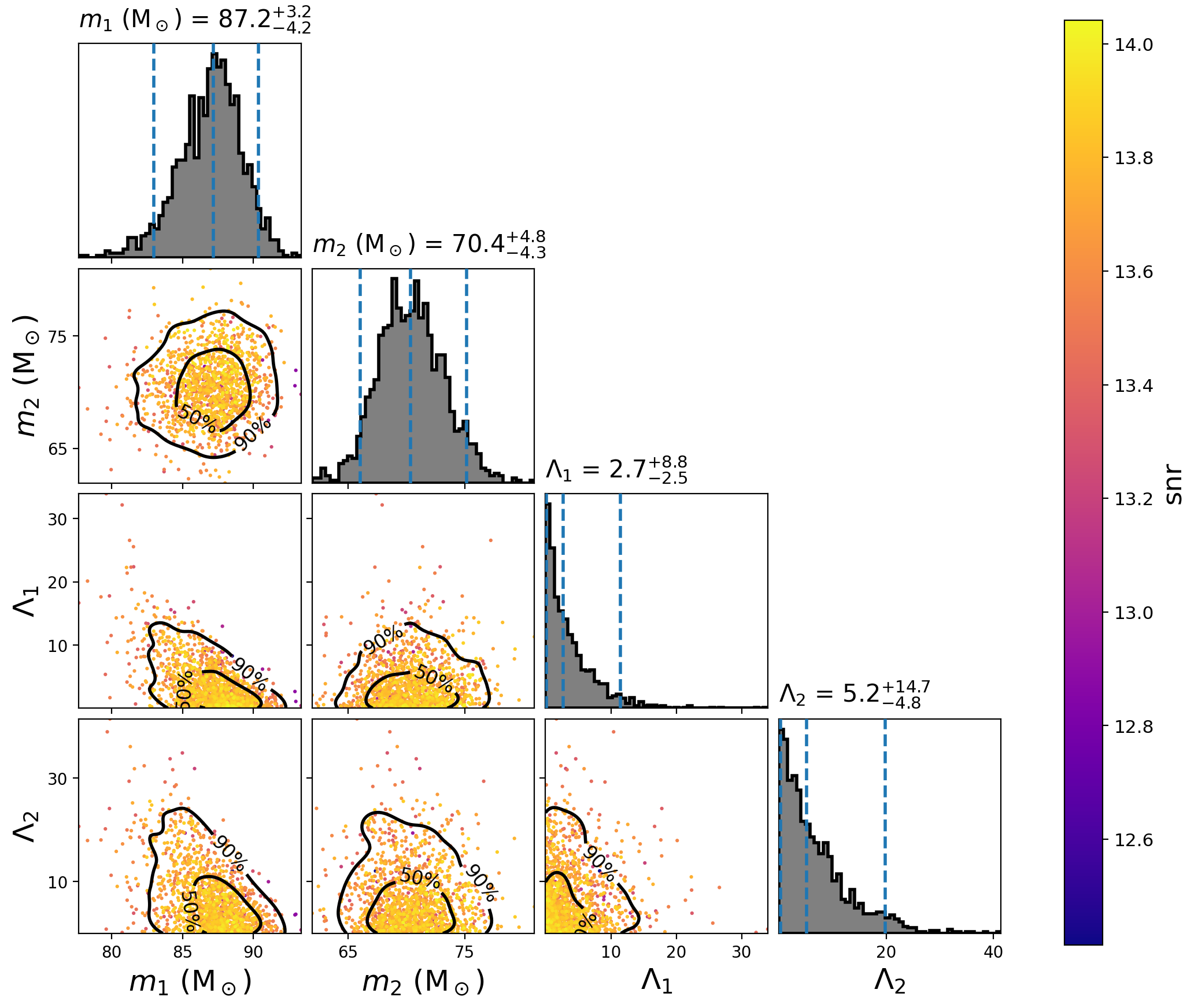}
    \caption{The GW190521 posterior probability density with IMRPhenomPv2\_NRTidal waveform template bank.}
    \label{fig:GW190521}
\end{figure}

We first consider GW190521  {and follow the standard procedure by LIGO and Virgo \cite{LIGOScientific:2021usb}}, the candidate event in the upper mass gap, and show the resultant posterior in Fig. \ref{fig:GW190521} by using the ``IMRPhenomPv2\_NRTidal'' waveform model for the matched filtering. Each colored point in Fig. \ref{fig:GW190521} is reached by the MCMC walker in the last iteration with the SNR close to the LIGO released data, $14.3^{+0.5}_{-0.4}$. The results show that the TLN in each star is $\Lambda_1 = 2.7^{+8.8}_{-2.5}$ and $\Lambda_2 = 5.2^{+14.7}_{-4.8}$, both of which are lower than the values for the considered boson stars. The small values of TLN imply that the objects could be just BH or very rigid compact stars. The result is close to the boson star model BS $\phi^8$ prediction despite that.

The parameter estimation result of this event has been updated several times by LIGO Scientific Collaboration \cite{LIGOopendata}. The most recent one yields the masses to be $98.4^{+33.6}_{-21.7} M_\odot$ and $57.2^{+27.1}_{-30.1} M_\odot$ as given in GWTC-2.1 \cite{LIGOScientific:2021usb}. { However, our result is closer to the one given in GWTC-2.0 \cite{GWTC-2} with secondary mass $69.0^{+22.7}_{-23.1} M_\odot$. Our result shows little change due to the inclusion of TLN, and the SNR obtained is close to the parameter estimation results using "IMRPhenomPv3HM"\cite{khan;2020prd} as the waveform template released by LIGO, $14.22^{+0.30}_{-0.34}$.}

{  There may be a discrepancy in the results for the primary mass compared to those released by LIGO. One possible reason is that the waveform model used is only designed to model the inspiral of NS, not the complicated merger phase. Hence, it tapers the amplitude before the merger. Thus, it is unsuitable for analyzing this event, where the merger-ringdown produces most of the SNR. "IMRPhenomPv2\_NRTidal" is also specifically tuned to numerical relativity simulations of BNS and includes EOS-dependent effects in the tidal contribution. Therefore, it is not suitable for analyzing binary boson stars. However, we want to assume possible scenarios for parameter estimation results when searching for BH mimickers. Currently, we are using a BNS waveform template for a simple demonstration, hoping that better waveform templates will be available for improved analysis of BH mimickers in the future.}

\subsubsection{Lower Mass Gap: GW190814}

\begin{figure}[H]
    \centering
    \includegraphics[width=0.45\textwidth]{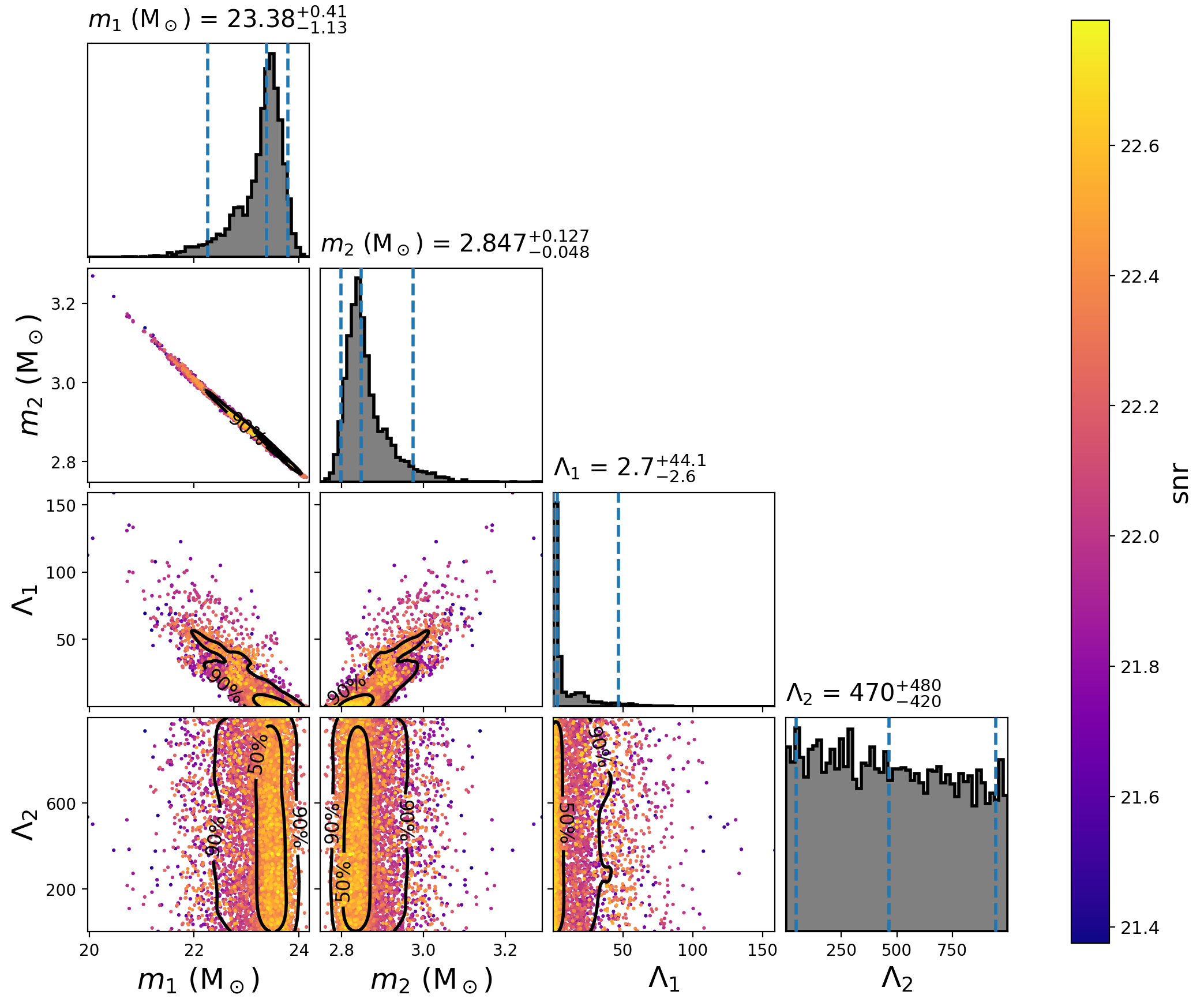}
    \caption{The GW190814 posterior probability density with IMRPhenomPv2\_NRTidal waveform template bank.}
    \label{fig:GW190814}
\end{figure}

The second example is the BH mimicker event in the lower mass gap, namely, the GW190814. The posterior is shown in Fig. \ref{fig:GW190814}, again with the "IMRPhenomPv2\_NRTidal" waveform model for the matched filtering. The SNR obtained { is close to the parameter estimation results using "IMRPhenomNSBH"\cite{Thompson;2020prd} as the waveform template released by LIGO, $23.41^{+1.65}_{-1.65}$}. We focus on the mass-gap component of the binary, the second one, which could be the BH mimicker. However, we obtain an almost flat posterior for $\Lambda_2$ with a given uniform prior in  $[0, 1000]$. Again, we cannot conclude whether this is a boson star based on our result. Our result is similar to the LIGO released properties of GW190814 \cite{LIGOScientific:2020zkf} based on the NSBH model with a uniform prior of $\Lambda_2 \in [0, 3000]$. The absence of measurable tidal features could be due to the highly unequal mass ratios and relatively large  {primary} masses, so the neutron star is tidally disrupted before the merger. Therefore, the resultant waveform shows little information about the tidal deformability of the massive secondary.

\subsection{Boson stars discrimination}
\paragraph{Tidal deformation in waveform---} 
The relevant difference in waveforms for boson stars, NS, and BH is mainly due to their associated TLNs. Especially, the TLN of BH is known to be zero\footnote{In case there is some cloud around BH due to phenomena like accretion or superradiance,  the TLN would deviate from zero,  as investigated in \cite{DeLuca:2021ite,  DeLuca:2022xlz}.} \cite{Charalambous:2022rre, Hui:2022vbh, Hui:2021vcv}. When considering the waveform of inspiral, the TLNs of both component objects will contribute to the waveform  {at the leading-order (5PN) correction through the combination \cite{Flanagan:2007ix,Favata:2013rwa,Wade:2014vqa} }
\begin{align}\label{Eq:effective Lambda}
\tilde{\Lambda} = \frac{16}{13}\frac{(m_1+12 m_2) m_1^4 \Lambda_1+(m_2+12 m_1) m_2^4 \Lambda_2}{M^5}.
\end{align} 
Later, we will focus on this quantity to distinguish boson stars from NS and BH through data analysis; see also \cite{Maselli_2017, Sennett:2017etc}.

\bigskip 

\paragraph{Bayesian model---} We would like to discriminate the boson stars from BH in a binary system, which is followed by two simple reasons: (i) most of the binaries observed by LIGO-Virgo-KAGRA (LVK) are BNS; 
and (ii) within our knowledge of boson stars, 
their tidal deformability is always tiny.

Based on Bayes’ statistics, we can perform hypothesis testing or model comparison by evaluating the Bayes factor. 
We take the hypothesis of BBH as the null hypothesis denoted by $H_0$, a binary boson star (BBS) and a boson star-black hole (BSBH) binaries are denoted by $H_1$ and $H_2$, respectively.  Without prejudice over the three hypotheses, we assume the same prior probability for them, i.e., $p(H_a) = 1/3$. 

The GW data analysis aims to extract the posteriors of the model parameters of the sources. There are intrinsic parameters, such as the masses and TLNs of the compact binary, and extrinsic parameters related to the location and orientation of the source relative to the detector. For simplicity, we will neglect the extrinsic parameters for a given strain data $d$ but characterize them by the signal-to-noise ratio (SNR) $\rho^{(d)}$ extracted from the matched filtering. For the intrinsic parameters, we will only consider the masses $m^{(d)}_{1,2}$ and TLNs $\Lambda^{(d)}_{1,2}$ of the binary. For BH, TLN is zero. On the other hand, for the boson stars, the TLN is related to the mass by the TLN-Mass relation, which can be denoted by  $\Lambda = \Lambda\big(m, b\big)$ with $b$ the model parameter of the associated EoS for SIDM. This implies that the combined TLN of \eq{Eq:effective Lambda} appearing in the waveform will be just the function of the binary masses $m_{1,2}$ and $b$, or denoted by 
\be
\tilde{\Lambda} = \tilde{\Lambda}(m_1, m_2, b)\;.
\ee
Since $\tilde{\Lambda}$ is crucial for the current hypothesis testing, we will then use it to construct the Bayesian inference.

Furthermore, in mocking data analysis, we need to generate the mocking strain data set $d_k$ for $k$-th GW event by superposing the corresponding theoretical waveform with real-time detector noise generated from the power spectral density. We then perform Bayesian inference using MCMC or nested sampling algorithm to scan the likelihood. Since the parameter estimation based on MCMC is very time-consuming, as a proof-of-concept study, we assume the Gaussian likelihood, i.e.,
\begin{align}\label{likelihood_f}
p (d_k|\tilde{\Lambda}, H_a ) 
&= \mathcal{N}\Big( \tilde{\Lambda}(d_k); \tilde{\Lambda}^{(th)}(H_a), \sigma^{(d)}_{\tilde{\Lambda}}(d_k) \Big),  \nonumber\\
&= \frac{1}{\sqrt{2\pi}\sigma^{(d)}_{\tilde{\Lambda}}(d_k )} 
\exp\Bigg[
  -\frac{ \big(\tilde{\Lambda}(d_k) - \tilde{\Lambda}^{(th)}(H_a) \big)^2 }
  {2\, \big(\sigma^{(d)}_{\tilde{\Lambda}}(d_k)\big)^2} \Bigg].
\end{align}
In the above,  {the theoretical value of the effective TLN is represented by $\tilde{\Lambda}^{(th)}(H_a)$ which should be calculated through the model of EoS and determined by prior, and} the "standard deviation" $\sigma^{(d)}_{\tilde{\Lambda}}(d_k )$ should be inherited from the detector noise so that a louder event, i.e., large SNR, has a smaller deviation. Assuming the universal detector noise, it can then be gauged by the known GW events. In our case, we choose GW170817 as the reference event so that 
\begin{align}
\sigma_{\tilde{\Lambda}}(\rho) &= \frac{\rho^{GW170817}}{\rho}\sigma_{\tilde{\Lambda}}^{GW170817},
\end{align}
where SNR $\rho$ obeys a universal power-law distribution,
\cite{Schutz:2011tw}
\begin{align}
f(\rho) = \frac{3\rho^{3}_{th}}{\rho^4},
\end{align}
where the threshold SNR is taken to be $\rho_{th} =12$.

Therefore, in our Bayesian model, we consider 
the effective tidal deformability $\tilde{\Lambda}$ as the key random variable, whose prior $p(\tilde{\Lambda} | H_a)$ is determined from the priors of  
$\{m_1, m_2, b_a\}$ from the TLN-Mass relation of a given SIDM model. For simplicity, we fix $b_a=b$ and assume flat priors in a finite range for the parameters $\{m_1, m_2\}$, i.e.,
\begin{subequations}
\begin{align}
p_a(m_1) &= \mathrm{Uniform}(m_1 | m_{1,\text{lower}}, m_{1,\text{upper}}, H_{a} ), \\
p_a(m_2) &= \mathrm{Uniform}(m_2 |  m_{2,\text{lower}}, m_{2,\text{upper}}, H_{a} ).
\end{align}
\end{subequations}
The $m_{1,2; {\rm lower/upper}}$ could be hypothesis dependent. We will examine the effect of EoS parameter $b$. 

With the above preparation for the Bayesian inference for the mock data analysis, for a given mock strain data $d$, the Bayes factor for the hypothesis testing of discriminating boson stars from the BH is
\be\label{Eq:bayes_factor_1}
\mathcal{B}_a = \frac{\int d\tilde{\Lambda}\, p(\tilde{\Lambda} | H_a) p(d| \tilde{\Lambda},H_a)}{
 p(d |\tilde{\Lambda}=0,H_0)}\;, \qquad a=1,2.
\ee
Since the mock data $d$ should be associated with a given set of binary masses and EoS parameter, i.e., $d=d(m_{1,2}^{(d)}, b^{(d)}, \rho^{(d)})$. Therefore, the resultant Bayes factor should also be a function of these parameters, i.e.,
\be
B_a(d)=B_a(m_{1,2}^{(d)}, b^{(d)}, \rho^{(d)})\;.
\ee
The tomography of $B_a$ encodes the population distribution of boson stars. Especially for the events in the mass gap, there is no primary BBH event. One can infer the boson star population from the Bayes factor, which reflects the underlying TLN-Mass relation. This tomography can be verified or falsified with future observation data on the gap events. Finally, we should comment that the methodology presented in the section is to examine the probability of the BH mimickers located in gap events based on Bayesian analysis. In other words, we tend to provide a way to determine if the gap events may look like certain ECOs rather than ordinary BH. If we wish to answer questions such as the composition of the ECO, one would need numerical relativity to generate the specific waveforms for a chosen DM model. One should be able to calibrate the inspiral-merger-ringdown waveforms, particularly in the duration of the merger. With such waveform models for DS, hopefully, we can provide the answer to the nature of DM.

\section{Conclusion}\label{conclusion}

In this review, we have discussed various properties of DSs, mainly the BS supported by self-interactions motivated by particle physics and string theory. Moreover, we have focused on the implication of these BS candidates to the GW observations. We believe the DS will be our universe's universal component if nontrivial DM exists with self-interactions. By the nature of DM, GW emitted from binary DSs will be the most natural and reliable way to detect them and explore their properties. We hope our review can serve as an introductory summary of this topic and guide the novices with an overview of the relevant issues. The DS is a topic covering a wide range of knowledge, and our review can't be complete. We did not discuss much on the formation issues but just some very preliminary proposals. A full-scale study of the DS formations should involve the N-body simulation based on some particular SIDM model and should wait for future works. To be honest, the main obstacle to answering the above difficulties lies in the nature of DM, namely, whether the DM is a single kind of particle or purely modified gravity as the explanation for the missing energy. The mass scale of DM, a fundamental particle or a composite one, the forms and strengths of DM to ordinary particles and themselves. All those properties would change the complete picture of DS formation and populations. Despite that, we hope our review can motivate extensive research on the DS and their observations. 

\bigskip
\bigskip

\noindent {\it Acknowledgements.} The authors thank Alessandro Parisi and Jing-Yi Wu for helpful discussions.
FLL and LWL are supported by the National Science and Technology Council (Taiwan), Grant 109-2112-M- 003-007-MY3. CSC is supported by the National Science and Technology Council (Taiwan), Grant MOST 110-2112-M-032-008-. KZ (Hong Zhang) thanks Shanghai City for the fund, which unfortunately cannot be disclosed here.

\appendix
\section{Derivation of isotropic Equations of State}
\label{apA}
In this paper, we study the canonical (massive) complex scalar for boson stars with the following Lagrangian
\be\label{effective}
{\cal L}=-{1\over 2} g^{\mu\nu} \partial_{\mu}\phi^* \partial_{\nu} \phi - V(|\phi|),
\ee
where
\be
V(\phi)={1\over 2}m^2 |\phi|^2 + U(|\phi|)
\ee
where $U$ is the potential for the self-interactions.

 Under spherical symmetry, we consider the 
 following the metric ansatz form
\ba\label{metrica_a}
ds^2=-B(r) d t^2+A(r) d r^2 + r^2 d \Omega\;.
\ea
In such a background, we assume the following stationary and spherical symmetric ansatz for the scalar field,
\be 
\phi(t,r)=e^{-i \omega t} \Phi(r)\;.
\ee

The field equations for the boson star configurations contain two parts: The first part is the Einstein equation with anisotropic stress tensor given by $T^{\mu}_{\nu}={\rm diag}(-\rho, p, p_{\perp}, p_{\perp})$ with 
\bea
\rho &=& m^2 M_{\rm pl}^2 \Big[{1\over 2}({\Omega^2 \over B}+1)\sigma^2 +{|\partial_x\sigma|^2 \over 2 A}+{U(M_{\rm pl} \sigma)\over m^2 M^2_{\rm pl}} \Big] \label{rho_aniso} \qquad \\
p &=& m^2 M_{\rm pl}^2 \Big[{1\over 2}({\Omega^2 \over B}-1)\sigma^2 +{|\partial_x\sigma|^2 \over 2 A}-{U(M_{\rm pl} \sigma)\over m^2 M^2_{\rm pl}} \Big] \label{p_aniso}\\
p_{\perp} &=& p - {m^2 M_{\rm pl}^2 |\partial_x\sigma|^2 \over  A}\;, 
\eea
where we have introduced the following dimensionless scaled quantities
\be
x=m r\;, \qquad \Omega={\omega \over m}\;, \qquad \sigma={\Phi \over M_{\rm pl}}
\ee
with the Planckian mass $M_{\rm pl}=1/\sqrt{4\pi G_N}$. Note that the anisotropy of the pressures is due to the gradient term $|\partial_x \sigma|^2$. 
The second part of the field equations is the one for $\Phi$ (or now $\sigma$),
\be\label{scalar_eq}
\partial_x(x^2 \sqrt{B\over A} \partial_x \sigma) + x^2 \sqrt{AB}\Big[({\Omega^2 \over B}-1)\sigma - {U'(M_{\rm pl} \sigma)\over m^2 M_{\rm pl}} \Big] =0\,,
\ee
where $U'(M_{\rm pl} \sigma):={\delta U(X) \over \delta X}|_{X=M_{\rm pl} \sigma}$.

Due to the anisotropic stress tensor, the Einstein equation cannot be further reduced to TOV equations. The only way to solve the boson star configuration is to solve the coupled field equations by the shooting method, which is numerically difficult to obtain the full mass-radius or even mass-TLN relations. One simplification is to assume the scalar field profile is almost flat inside the boson stars so that the gradient term $\partial_x\sigma|^2$ in \eq{rho_aniso}-\eq{p_aniso} and \eq{scalar_eq} can be dropped. However, the gradient term cannot always be small unless for some particular self-interaction regime. To see this, let us assume that for some proper choice of a dimensionless parameter $\Lambda$, which is the combination of $m$, $\sigma$, and coupling constants in $U$, one can have
\be\label{iso-c}
{U(M_{\rm pl} \sigma)\over m^2 M^2_{\rm pl}} \longrightarrow {U_*(\sigma_*)\over {\Lambda}}\qquad \mbox{where} \quad  \sigma_*=\sqrt{\Lambda} \sigma
\ee
with $U_*$ being in a universal form involving none of the model parameters in $V$. Note that this above implies
\be
{U'(M_{pl} \sigma)\over m^2 M_{\rm pl}}\rightarrow {U'_*(\sigma_*)\over \sqrt{\Lambda}} \qquad \mbox{with} \quad  U'_*(X)={\delta U_*(X) \over \delta X}.
\ee
If we further scale $x$ by $x_*=x/\sqrt{\Lambda}$ and assume that $A$ and $B$ are not affected by this scaling, then it is straightforward to see that the terms involving the gradient $\partial_x \sigma$ in \eq{rho_aniso}-\eq{p_aniso} and \eq{scalar_eq} are suppressed by $1/\Lambda$ factor than the other non-gradient terms in the corresponding equations. Therefore, the gradient terms can be neglected in the large $\Lambda$ limit. In this case, scalar field equation \eq{scalar_eq} is reduced to the algebraic one
\bea\label{scalar-c}
({\Omega^2 \over B}-1)\sigma_* = U'_*(\sigma_*)\;.
\eea 
Then we obtain that, \eqref{rho_aniso} and \eqref{p_aniso}  reduce to 
\bea
\rho &=& \rho_0 \Big[{1\over 2} \sigma_* U'_*(\sigma_*) + U_*(\sigma_*) +\sigma_*^2 \Big]\;, \label{rho-fn} \\
p&=&p_{\perp} = \rho_0 \Big[{1\over 2} \sigma_* U'_*(\sigma_*) - U_*(\sigma_*) \Big]\;, \label{p-fn}
\eea 
and the overall energy density scale
\be \label{rho0}
\rho_0 ={ m^2 M_{\rm pl}^2 \over \Lambda}\;.
\ee
Then, for specific EoS, we can insert the corresponding potential $U_*$ and find the results listed in the main text.

\section{Bondi accretion for non-relativistic and relativistic fluids}\label{Bondi_accr}
Bondi accretion \cite{Bondi:1944jm,Bondi:1952ni} is the simplest accretion scenario by assuming spherical symmetry.
It can be the approximate mechanism to create the boson stars or the DM spike around a central BH. 
Consider first the non-relativistic fluid, i.e., the sound speed is far less than the speed of light. Kinematically, the continuity equation ${1\over r^2} {d\over dr}(r^2 \rho_0 v)=0$ defines the accretion rate 
\be\label{continuity}
\dot{M}= 4\pi \rho_0 r^2 v
\ee
where $M$ is the total mass of the central object,  { $\rho_0$ is the mass} density of the fluid, $r$ is the radial distance, and $v$ is the  { magnitude of the inward} velocity of the fluid element. On the other hand, the dynamics are dictated by the Euler equation
\be 
v {d v \over dr} + {1\over \rho_0}{d p \over dr}+{G_N M \over r^2}=0
\ee
where $p$ is the pressure of the fluid, and $G_N$ is the Newton constant. Use the definition of the sound speed $c_s^2={d p\over d\rho_0}$, we can turn the Euler equation into
\be\label{Bernoulli}
{1\over 2}\Big(1-{c_s^2 \over v^2}\Big) {d v^2\over dr}=-{G_N M\over r^2}\Big( 1-{2 c_s^2 r \over G_N M} \Big)\;.
\ee 
Note that $c_s=c_s(r)$. This equation predicts a sonic horizon at $r=r_s$ with $r_s={G_N M \over 2 c_s^2(r_s)}$, at which $v(r_s)=c_s(r_s)$. It is naturally assumed the fluid speed $v$ approaches zero at $r=0$, so the Euler equation tells that $v$ is monotonically increasing toward the sonic horizon; thus $v>c_s$ for $r<r_s$. If we further assume the equation of state of the fluid is in the polytropic form $p=\kappa \rho_0^{\gamma}$, the Euler equation yields
\be 
{v^2 \over 2}+{c_s^2 \over \gamma-1}-{G_N M \over r}=\mbox{constant}={c_s^2(\infty) \over \gamma-1}\;.
\ee 
Along with the above boundary condition, we obtain
\be 
c_s(r_s)=c_s(\infty) \Big( {2\over 5 -3 \gamma} \Big)^{1/2}\;. 
\ee 

On the other hand, from the continuity equation \eq{continuity}, we have $\dot{M}=-4\pi r^2 \rho_0 v=4\pi r_s^2 \rho_0(r_s) c_s(r_s)$, which can yield
\be
\rho_0(r_s)=\rho_0(\infty)\Big[{c_s(r_s)\over c_s(\infty)}\Big]^{2/(\gamma-1)}
\ee 
and 
\be \label{non_rel_accr_rate}
\dot{M}=\pi G_N^2 M^2 {\rho_0(\infty)\over c_s^3(\infty)}\Big[{2 \over 5 -3\gamma} \Big]^{(5-3\gamma)\over 2 (\gamma-1)}\;.
\ee 
Therefore, the accretion rate is determined by $\gamma$, the total mass $M$ of the central object, and the boundary values $\rho_0(\infty)$ and $c_s(\infty)$. 

However, the DM can be relativistic, e.g., the self-interacting relativistic bosonic scalar, then the above non-relativistic formulation of Bondi accretion can fail.  For the relativistic fluid, the energy density $\rho$ is different from the mass density $\rho_0$ used above, and they are related by \cite{Shapiro:1983du}
\be 
\Big({d \rho \over d \rho_0} \Big)_{\rm ad}={\rho + p \over \rho_0}
\ee 
where the subscript $\mbox{ad}$ indicates the variation is adiabatic. The continuity equation takes the same form as \eq{continuity}, however, the Bernoulli equation \eq{Bernoulli} should be replaced by its relativistic partner \cite{Shapiro:1983du,Richards:2021zbr,Feng:2021qkj} in the background Schwarzschild metric,
\be
\Big({p+\rho \over \rho_0} \Big)^2\Big( 1 + v^2 - {2 G_N M \over r} \Big) = \Big({p(\infty) +\rho(\infty) \over \rho_0(\infty)} \Big)^2\;,
\ee 
and the sonic horizon $r_s$ is at where the local Mach number equal to one \cite{Shapiro:1983du}, i.e., 
\be  
{v/c_s \over \sqrt{1-2G_N M/r_s + v^2}}=1\;.
\ee 
Thus, given the equation of state $p=p(\rho)$, we can solve the accretion rate $\dot{M}$ and the sound speed profile $c_s(r)$ in a similar way to the non-relativistic case. In \cite{Feng:2021qkj}, the relativistic Bondi accretion for the $\lambda |\phi^4|$ self-interaction bosonic scalar of mass $m$ has been studied. In this case, the accretion rate can be obtained analytically as follows,
\be  
\dot{M}\equiv 8\pi G_N^2 M^2 {\rho_B \over c^3} \Big({ c^2 + 3 c_s^2 \over c^2 - 3 c_s^2}  \Big)^{3/2} \sqrt{c^2 - c_s^2 \over c_s^2}\;,
\ee 
where $c$ is the speed of light and the normalization for the mass density
\be
\rho_B={3.48 \times 10^{20}\over \lambda} \Big({m\over \textrm{GeV}}\Big)^4 \textrm{kg}\; \textrm{m}^{-3}\;,
\ee 
and the sound speed in terms of the boundary value,
\be
c_s^2={c^2-3 c_s^2(\infty)+\sqrt{c^2 +66 c_s^2(\infty)-63 c_s^4(\infty)} \over 18(c^2-c_s^2(\infty))}\;.
\ee
Furthermore, note that $1/9\le c_s^2 < 1/3$ so that 
\be 
\dot{M}_{min} \le \dot{M} < \infty
\ee
where
\be\label{accr_rate_rel}
\dot{M}_{min}:= 64 \pi {G_N^2 M^2\over c^3}\rho_B\;. 
\ee

For a typical dark halo with a dispersion velocity of about $100 \;{\rm km} \;{\rm s}^{-1}$, 
\be 
\dot{M}_{min}\approx 1.4 \times  10^{-9}\Big({M \over \rm{M}_{\odot}} \Big)^2   \; M_{\odot} \; \rm{yr}^{-1}. 
\ee
This is too slow an accretion rate to form astrophysical-sized DS. Therefore, forming DS from pure Bondi accretion mechanism is hard. However, it can be adapted to consider the dark spike around a supermassive BH as considered in \cite{Feng:2021qkj}.  An interesting extension to the Bondi accretion around a moving BH can be found in \cite{Mach:2021zqe}.

\bibliographystyle{unsrt}
\bibliography{Review_v2.bib}

\end{document}